\documentclass[num-refs]{article}
\usepackage[margin=1.0in]{geometry}
\usepackage{setspace} 

\usepackage{harvard}
\usepackage{amsmath}
\usepackage{amssymb}
\usepackage{bbm}
\usepackage{epsfig}
\usepackage{subfigure}
\usepackage{arydshln} 
\usepackage{mathrsfs}

\newcommand{\bx}{\boldsymbol{x}}

\newcommand{\bI}{\boldsymbol{I}}
\newcommand{\bbeta}{\boldsymbol{\beta}}
\newcommand{\bmu}{\boldsymbol{\mu}}

\newcommand{\Xcur}{\mathscr{X}}
\newcommand*\rot{\rotatebox{90}}

\title{A Copula-based Fully Bayesian Nonparametric Evaluation of Cardiovascular Risk Markers for Normoglycemic Patients in the Mexico City Diabetes Study}

\author{Claudia Wehrhahn\footnote{Department of Statistics, University of California at  Santa Cruz, Santa Cruz, CA, 95064, USA; cwehrhah@ucsc.edu}, 
Ruth Fuentes-Garc\'ia\footnote{Department of Probability and Statistics,  Universidad Nacional Aut\'onoma de M\'exico, Ciudad de M\'exico, 06320,  M\'exico}, 
Rams\'es H. Mena\footnote{Department of Probability and Statistics,  Universidad Nacional Aut\'onoma de M\'exico, Ciudad de M\'exico, 06320,  M\'exico},  
Fabrizio Leisen\footnote{School of Mathematics, Statistics and Actuarial Science, University of Kent, Canterbury, CT2 7NZ, UK}, \\Maria Elena Gonz\'alez-Villalpando\footnote{Center for Studies in Diabetes A.C., Ciudad de M\'exico, 06320 , M\'exico}, and Clicerio Gonz\'alez-Villalpando\footnote{Center for Studies in Diabetes A.C., Ciudad de M\'exico, 06320 , M\'exico, National Institute of Public Health, Morelos, 62100, M\'exico}}

\begin{document}

\maketitle

\begin{abstract}
Cardiovascular disease leads the cause of death worldwide and several studies have been carried out to understand and explore cardiovascular risk markers in normoglycemic and diabetic populations.  In this work, we explore  the association structure between hyperglycemic markers and cardiovascular risk markers controlled by triglycerides, body mass index, age and sex, for  the normoglycemic population in The Mexico City Diabetes Study. Understanding  the association structure  contributes to the assessment of additional cardiovascular risk markers in this low income urban population with a high prevalence of classic cardiovascular risk biomarkers.  The association structure is measured by conditional Kendall's tau, defined by means of conditional copula functions. The latter are in turn modeled under a fully Bayesian nonparametric approach, which allows the complete shape of the copula function to vary for different values of controlled covariates, which  mediate the association.
\end{abstract}


\section{Introduction} \label{sec:intro}

Considerable research has been carried out to understand the complex pathogenesis of Type 2 Diabetes (T2D) as well as its role as a cardiovascular risk factor in Mexico (\cite{glzetal2014},  \cite{escobedoetal2011}).  Although a higher risk of cardiovascular disease is expected for diabetic patients, this relation is not so clear for normoglycemic patients, i.e., patients with normal sugar blood levels. In normoglycemic patients the identification and  understanding of possible cardiovascular risk  markers is  important to monitor the potential development of cardiovascular disease (\cite{thomasetal2006}).  One question that has arisen in the literature is whether the glycated hemoglobin (HbA1c) test can provide more information about cardiovascular risk compared to that provided by the 2-hour postchallenge glucose (PG2-H) test. \cite{huang2011glycated} evaluated the association between hyperglycemic  and cardiovascular risk markers, which has been reported as non consistent for non-diabetic subjects, in a population from Shanghai. In particular,  using  linear regression models considering the quartiles of the glycemic markers and adjusting by sex, smoking and drinking status, body mass index (BMI), blood pressure, and serum lipids, they analysed the association  between HbA1c and PG2-H with the carotid  intima-media thickness  (CIMT) test, a  cardiovascular risk  marker. They reported that the population of normoglycemic subjects they analysed had a higher association between HbA1c and CIMT than  between PG2-H and CIMT, suggesting HbA1c could be more informative of cardiovascular risk.   To contribute to this discussion, our analysis aims to provide guidelines of possible association between cardiovascular and hyperglycemic risk markers,  for normoglycemic patients in the Mexico City Diabetes Study (MCDS).  The MCDS is a prospective, population-based, and clinico-epidemiological study about T2D and follows a low income cohort of elder urban participants in Mexico City.  We focus on the normoglycemic patients from the MCDS and our aim is to investigate the use of hyperglycemic markers such as PG2-H or HbA1c as cardiovascular risk markers, controlled by relevant covariates. Although our interest is not on causal relations, covariates such as triglycerides, BMI, age, and sex could mediate the association, therefore they are included in our study. 

We study the association structure between  blood sugar measurements 
 and cardiovascular risk markers by means of the Kendall rank correlation coefficient, hereafter named Kendall's tau, which measures the ordinal association between variables.  Specifically, we model Kendall's tau between PG2-H and CIMT and between HbA1c and CIMT for different values of covariates such as  triglycerides levels, BMI, age, and sex. The use of classical nonparametric rank correlation tests in the analysis of the MCDS dataset  results in non statistical significance, moreover it
 does not highlight  the dynamics of the association structure for different sets of covariates. An alternative approach is to model conditional copula functions which in turn define conditional Kendall's tau. This approach provides a complete characterization of the conditional association structure between cardiovascular and hyperglicemic markers, modulated by covariates. However, modeling conditional  copula functions under a parametric approach leads to a model selection problem. A more flexible modeling approach avoids this selection and provides a better understanding of the association structures. Therefore, we use a fully Bayesian nonparametric  approach to model the covariate dependent copula functions between PG2-H and CIMT and between HbA1c  and CIMT for varying values of covariates. Through a post processing analysis, we obtain the dynamics of the corresponding conditional Kendall's tau. 

Bayesian nonparametric (BNP) models have been widely used,  and their advantages, compared to their parametric counterparts, are abound. These models relax standard parametric assumptions, such as the choice of a particular family of distributions or the number of components that a mixture model requires. Due to its strengths, the use of  BNP models for copula function estimation has increased during the last years. Regarding unconditional copula functions,   \cite{wu2014bayesian} and \cite{wu2015bayesian} proposed  BNP models for  copula density estimation based on infinite mixtures of Gaussian and skew-normal copulas, respectively.  \cite{burda2014copula} modeled the dependency structure between random variables by first modeling the marginals as infinite mixtures, and then linking them through a random Bernstein polynomial copula function, while \cite{ning2018nonparametric} modeled the copula function using Dirichlet polya trees.  In the context of predictor-dependent copula function modeling, \cite{hernandez2013gaussian}, \cite{lopez2013gaussian}, and \cite{levi2018bayesian} considered transformations of Gaussian processes in the definition of the copula function. Although  these approaches are appealing, they require the selection of  a specific copula function, which leads to a model selection problem. \cite{leisen2017bayesian} extended the  model  proposed by \cite{wu2015bayesian} and defined an infinite mixture of conditional Gaussian  copula functions, which  is flexible, but relies on the selection of a calibration function that relates the correlation structure of the copula function with the predictors. Our modeling approach extends the model proposed by~\cite{leisen2017bayesian}  and overcomes the before mentioned drawbacks.

Here, we model conditional copula functions under a fully BNP approach.  Under this modeling framework  a prior distribution is placed on sets of predictor-dependent random measures involved in the definition of the model.  More specifically, our fully BNP approach considers a predictor-dependent stick-breaking prior distribution for the collection of predictor-dependent random measure,  as proposed by \cite{maceachern;2000}. Under this framework, for each vector of covariates, the conditional copula function is modeled as an infinite mixture of copula functions, therefore  copula functions flexibly vary across covariates and a flexible dynamic characterization of the association structure for varying values of the covariates is obtained. As will be shown later, the borrowing of information between covariates and flexibility of  such a modeling approach provides improvements compared to the modeling framework proposed by~\cite{leisen2017bayesian}.  Fully BNP models based on the dependent stick-breaking and other extensions have been proposed by \cite{deiorio;mueller;rosner;maceachern;2004}, \cite{gelfand;kottas;maceachern;2005}, \cite{deiorio;johnson;mueller;rosner;2009}, \cite{rodriguez2011nonparametric},   \cite{barrientos2017fully}, among others.

The rest of the manuscript is organized as follows. Section~\ref{sec:MCDS} describes the MCDS and motivates the analysis for the real data set while Section~\ref{sec:meth} provides background and the description of the model.   Section~\ref{sec:simulation} illustrates the performance of the model in  synthetic data sets while Section~\ref{sec:appl} describes the association analysis for the MCDS. Finally, Section~\ref{sec:conc} concludes the manuscript.

\section{The Mexico City Diabetes Study} \label{sec:MCDS}
 
The MCDS (\cite{gonzalez1999mexico}) study began in 1989 with the identification of a homogeneous low-income site. All men and non pregnant women aged 35 to 64 years were defined as eligible. The research protocol, informed consent, procedures, and methods were approved by the Institutional Review Board of the Center for Studies in Diabetes and all participants gave informed consent.

At baseline (1989 to 1990), a total of $3,319$ participants were interviewed and $2,282$ examined (from a total of $3,505$ eligible individuals). The final cohort (interviewed and examined) was composed of $941$ men and $1,341$ women. During the last phase (2008 to 2009), for a total of $1,174$ participants,  in addition to the standardized exam and oral glucose tolerance test, an HbA1c measurement was included. T2D was diagnosed using American Diabetes Association criteria: fasting plasma glucose concentration $126$ mg/dL and/or a 2-hour plasma glucose concentration $200$ mg/dL after a standard 75-g glucose load. Participants who self-reported a history of diabetes and were taking oral glucose-lowering agents were considered to have T2D, regardless of their plasma glucose values. Pre-diabetes  (PD)  was diagnosed when an individual had a fasting plasma glucose of $100$ to $125$ mg/dL and/or a 2-hour postglucose load between $140$ and $199$ mg/dL. 

In our analysis, we consider the subset of normoglycemic individuals from the MCDS at the last phase of the study who had the CIMT measured in phase 3 and who had measurements for all the considered covariates.  Since the CIMT is prone to measurement error,  values lower than the first quartile and greater than the third quartile were excluded. The resulting data sets contain $n=246$ individuals. As covariates we include triglycerides, BMI, sex  and age, the latter discretized into four categories. Table~\ref{tab:desc2} provides summaries for the variables involved in the analysis.  The median age of the participants is $59$ years and $58.9$\% are women.  Although our analysis focuses on the normoglycemic subset from the MCDS  we note that more than $84$\% of this group shows overweight or obesity ($BMI > 25$) and more than $42$\% has borderline, high, or very high triglycerides levels ($Triglycerides > 150$).

\begin{table}
\caption{\label{tab:desc2} Descriptive statistics for  variables involved in the analysis. Summaries for continuous variables are presented as median (interquartile range) while  discrete variables are presented as  count  $( \%)$.}
\centering
\begin{tabular}{lc|lc|lc}
\multicolumn{2}{c}{Continuous Marker}  & \multicolumn{2}{c}{Continuous Covariate}   & \multicolumn{2}{c}{Discrete Covariate}   \\ \hline
HbA1c  &  6.3 (1.1)  & Triglycerides         & 141.5 (91) &  Female  sex & 145 (58.9 \%)\\
PG2-H & 102.5 (30)  & BMI           &   28.3  (4.98)&  $55 \leq Age  < 60 $ & 78 (31.71 \%) \\
CIMT & 0.64  (0.15)  & & & $60 \leq Age  < 65$ &  49 (19.92 \%) \\
 & & & & $Age  \geq 65$ &  66 (26.83 \%) \\  
\hline 
\end{tabular}
\end{table}

To contribute in the assessment of potential use of HbA1c as a cardiovascular risk marker, we explore the association between HbA1c and  CIMT, controlled by  triglycerides, BMI, age, and sex, and compare these results with those obtained from analyzing the association between  PG2-H and CIMT, controlled by the same set of covariates. Our interest is not on causal relations between HbA1c and  CIMT or PG2-H and CIMT. As a matter of fact, if we were to fit  a linear regression model for CIMT with  HbA1c and  PG2-H  as independent variables none of them would  be statistically significant and the inclusion of additional covariates would not change these results.

\section{Background and Methods}
\label{sec:meth}

It is well known that copula functions, introduced  in the statistical literature by \cite{sklar1959fonctions},  describe the dependency structure between two or more random variables.  Roughly speaking, copulas link the univariate marginal distribution function of random variables  with their joint cumulative distribution function (c.d.f.).  Since our interest is to model the dependency structure between pairs of random variables, in what follows we give a brief description of two-dimensional copulas. 

A two-dimensional copula function, denoted $C$,  is a c.d.f. on $[0,1]^2$ such that $C(u, 0) = C(0, u) = 0$, $C(u, 1)=C(1, u) = u$, for every $u\in[0,1]$, and $C(A(u_1, u_2)) \geq 0 $ for every set $A(u_1, u_2) \subseteq [0,1]^2$.  According to Sklar's theorem, the joint c.d.f., denoted $H$, of two random variables, say $Y_1$ and $Y_2$, with  univariate marginals $F_1$ and $F_2$, respectively, can be linked through a copula function, which is unique, when $F_1$ and $F_2$ are continuous. The relation is as follows: $H(y_1, y_2) = C(F_1(y_1), F_2(y_2))$.  When the bivariate c.d.f. has a probability density function (p.d.f.), denoted $h$, it holds that $h(y_1, y_2)=c(F_1(y_1), F_2(y_2))f_{Y_1}(y_1)f_{Y_2}(y_2)$, where $c$ is the copula density and $f_{Y_1}$  and $f_{Y_2}$ are the p.d.f.s of $Y_1$ and $Y_2$, repectively. A formal definition of copula functions, their properties, and more, can be found in \cite{nelsen1999introduction}. 

Given the practicability for covariate inclusion, here we concentrate on the family of Gaussian copulas.   Let $\Phi(y_1, y_2 \mid \rho)$  denote the c.d.f. of a bivariate normal distribution with zero mean vector and  correlation parameter $\rho$ and let   $\Phi^{-1}(p)$ denote the inverse c.d.f. of a standard normal distribution,  $p\in(0,1)$. The Gaussian copula function, denoted $C^{G}(u_1, u_2\mid \rho)$,  of the bivariate vector $(u_1, u_2)\in [0,1]^2$, with correlation parameter $\rho\in (-1,1)$,   is given by $C^{G}(u_1, u_2\mid \rho):=\Phi(\Phi^{-1}(u_1), \Phi^{-1}(u_2)\mid \rho)$.  The associated Gaussian copula density, denoted $c^{G}(u_1, u_2\mid \rho)$, is given by
\begin{eqnarray}\label{GaussianCopula}
\!\!\!\!\!\!\!\!\!\!\!\!\! c^{G}(u_1, u_2\mid \rho)&   = &\vert \Sigma_{\rho} \vert^{-1/2}\exp\left\{-\frac{1}{2} (\Phi^{-1}(u_1), \Phi^{-1}(u_2))(\Sigma_{\rho}^{-1}-\bI_2)    {{\Phi^{-1}(u_1)}\choose{\Phi^{-1}(u_2)}} \right\}, 
\end{eqnarray}
where $\bI_2$ is the $2$-dimensional identity matrix and 
\[\Sigma_{\rho}=\begin{pmatrix}
    1       &\rho \\
   \rho       &1 \\
    \end{pmatrix}.\]

Copula functions play a key role in the study of the association between random variables.  In particular,  one scale-invariant  measure of association  is Kendall's tau, denoted $\tau$. Specifically,  Kendall's tau measures the concordance between random variables, is defined by means of copulas, and is given by 
\begin{align*}
\tau &= 4E\left[C(U_1, U_2)\right] -1 = 4\int_0^1\int_0^1  C(u_1, u_2) dC(u_1, u_2) - 1.
\end{align*}

Our interest is to model the dependence structure between cardiovascular  and hyperglycemic risk markers, controlled by relevant variables. Thus,  we need the dependence  structure to be dynamic for  varying values of such variables. There is a rich literature on conditional copula functions and measures of association. See \cite{lambert2002copula},  \cite{patton2006modelling},  \cite{hafner2010efficient},  \cite{gijbels2011conditional}, \cite{veraverbeke2011estimation},   \cite{acar2011dependence}, \cite{gijbels2012multivariate},  \cite{abegaz2012semiparametric},  among others. 

Conditional joint distributions are completely described by the conditional marginals and a conditional copula function, which is unique when the conditional marginals are continuous.  According to Sklar's theorem the relation is as follows $H(y_1, y_2 \mid \bx) = C(F_1(y_1 \mid \bx), F_2(y_2 \mid \bx)\mid {\bx})$, where  $C(\cdot \mid {\bx})$ is the  predictor dependent copula function that describes the dependence structure between two random variables for varying values of the predictor $\bx$. This, in turn, directly defines predictor dependent measures of association such as conditional Kendall's tau, given by  
\begin{align}\label{conditionalTau}
\tau (\bx)&= 4E\left[C(U_1, U_2\mid \bx)\right] -1 = 4\int_0^1\int_0^1  C(u_1, u_2\mid \bx) dC(u_1, u_2\mid \bx) -1.
\end{align}

As stated before, the  purpose here is to robustly  model predictor dependent copula functions to describe the predictor dependent association structure between cardiovascular  and hyperglycemic risk markers. Dependent BNP models provide a flexible  tool for such data modeling framework.  In what follows we describe our covariate adjusted copula function modeling approach and its use to measure the conditional  association structure between cardiovascular and hyperglycemic risk markers.

We consider a  Dirichlet process mixture  of  bivariate  Gaussian copula densities model with predictor dependent mixing distribution. Specifically, we use a dependent stick-breaking prior distribution for the set of predictor dependent random measures, as proposed by \cite{maceachern;2000}, who extended the well known stick-breaking process prior proposed by \cite{sethuraman;94} to accommodate for covariate dependency. The model can be written as
\begin{align}\label{DPMofCopulas}
c(u_1, u_2\mid \bx) &= \int c^{G}(u_1, u_2 \mid z)G_{\bx}(dz),  \quad\quad (u_1, u_2)\in [0,1]^2,
\end{align}
where $c^{G}$ is defined as in Equation (\ref{GaussianCopula}) and the set   $\{G_{\bx} : \bx \in \Xcur\}$ follows a dependent stick-breaking process  prior such that, for each $\bx\in\Xcur$, the predictor dependent random measure is given by  $G_{\bx}(\cdot)=\sum_{j=1}^{\infty}w_j(\bx)\delta_{\rho_j(\bx)}(\cdot)$, with $w_j(\bx)=v_j(\bx)\prod_{l<j}\left[1-v_l(\bx)\right]$, where the stick-breaking variables, $\{v_j(\bx)\}_{j=1}^{\infty}$,  and correlations, $\{\rho_j(\bx)\}_{j=1}^{\infty}$, are defined by transformations of  real--valued stochastic processes. Note that, both weights and correlations are dynamic. To define  $v_j(\bx)$ and $\rho_j(\bx)$ we exploit the relation between Gaussian processes and linear regression models, and consider suitable transformations. Note that the model in Equation (\ref{DPMofCopulas}) can be equivalently written as 
\begin{align}\label{SBCopulas}
c(u_1, u_2\mid \bx) &= \sum_{j=1}^{\infty}  w_j(\bx) \ c^{G}(u_1, u_2 \mid \rho_j(\bx)),  \quad\quad (u_1, u_2)\in [0,1]^2.
\end{align}

Under this approach the copula density  is flexibly modeled through mixtures of bivariate Gaussian copula densities that  vary and accommodate for different values of the predictor.  We consider $v_j(\bx)$   defined by the logistic transformation of the linear predictor, i.e., $v_j(\bx)=e^{\bx^t\bbeta^{v}_j}/(1+e^{\bx^t\bbeta^{v}_j})$, with $\bbeta^{v}_j\overset{iid}{\sim} N_p(\bmu^{v}, \Sigma^{v})$, $j\geq 1$. Similarly, the $\rho_j(\bx)$ processes are defined by $\rho_j(\bx)=2/(\vert \bx^t\bbeta^{\rho}_j\vert+1)-1$, with   $\bbeta^{\rho}_j\overset{iid}{\sim} N_p(\bmu^{\rho}, \Sigma^{\rho})$,  where $N_p(\bmu, \Sigma)$ denotes a $p$-dimensional normal distribution with parameters $\bmu\in \mathbb{R}^p$ and $\Sigma$  a $(p\times p)$-dimensional positive definite matrix. We termed this model as dependent Dirichlet process mixture of copula densities (DDPMC) model.  A sufficient and necessary   condition for the weights to add up to one in Equation (\ref{SBCopulas}) is that for every $\bx\in\Xcur$, $\sum_{j=1}^{\infty}\log\left[ 1-E(v_j(\bx))\right]=-\infty$ (Lemma 1 in \cite{ishwaran;james;2001}).


The conditional association structure between cardiovascular and hyperglycemic risk markers can be described by the covariate dependent  Kendall's tau, which as defined in Equation (\ref{conditionalTau}), depends on the dynamic copula functions. Under the DDPMC model inference for the association between  variables of interest is easily achieved for varying values of the predictors. 


\section{Model validation with simulated data}\label{sec:simulation}
We illustrate the performance of the model in two simulation scenarios that, to some extent, resemble the observed features  in the analysis of the MCDS data set. For each simulation scenario, one sample of size 250 was generated  from the bivariate joint distribution defined by a  copula function and standard normal marginals. In order to have data points that closely approximate samples $U_{i1}$ and $U_{i2}$ from the copula densities, we consider the pseudo observations $\tilde{U}_{i1} = R_{i1}/(n+1) $ and $\tilde{U}_{i2} = R_{i2} /(n+1)$, where $R_{il}$ denote  the rank of $Y_{il}$ among $Y_{1l}, \ldots, Y_{nl}$, $l=1,2$, and $(Y_{i1}, Y_{i2})$ denote the pairs generated from the joint distribution, as discussed by \cite{genest2009estimating}.

Under both scenarios, uniform on $(0,1)$ values were generated for the predictor. The first simulation scenario, named Scenario I, is defined by a predictor dependent elliptical $t$ copula with correlation parameter $\rho(x) = x$. Kendall's tau for Scenario I is an increasing  positive function  of the predictor that varies from 0.006 to 0.910 for values of the predictor between  0.010 and 0.990. The second simulation scenario, named Scenario II, is defined by a mixture of  two predictor dependent copula functions: a $t$ copula  with correlation parameter $\rho(x) = -x(1-x)^2$ and a Gumbel copula with parameter $\alpha(x) = x^2(1-x)+1$, and predictor dependent weights.  Under this scenario, Kendall's tau is an almost constant function of the predictor that varies between 0.0 and 0.015 showing a very low positive association between the random variables. For more details on the simulation scenarios, see the Supplementary Material.

We compare the performance of the proposed methodology with the model proposed by \cite{leisen2017bayesian}, which corresponds to a Dirichlet process mixture of conditional Gaussian copula densities model, here denoted LDVR model. Under this approach a stick-breaking process prior is assumed for a single random measure, which is fundamentally different from our proposal in which a dependent stick-breaking process prior is  assumed for the collection of covariate dependent random measures, allowing them to flexible vary for different values of the predictor. Additionally,  under the LDVR  model, the dependency on covariates is driven by the calibration function, denoted $\theta(\bx\mid \bbeta)$, $\bbeta\in \mathbb{R}^p$, for which a specific choice is required. The LDVR model is given by
\begin{align*}
c(u_1, u_2\mid \bx) &= \sum_{j=1}^{\infty}  w_j \ c^{G}(u_1, u_2 \mid \rho_j(\bx)),  
\end{align*}
where $c^{G}$ is defined as in Equation (\ref{GaussianCopula}), $\rho_j(\bx) = 2/(\vert \theta(\bx\mid \bbeta)\vert + 1) -1$,  and $w_j=v_j\prod_{l<j}(1-v_l)$, where $\bbeta \overset{iid}{\sim} N_p(\bmu, \Sigma)$,  $v_j\overset{iid}{\sim} Beta(1, \alpha)$, and $\alpha>0$. Regarding the calibration function we consider $\theta(\bx\mid \bbeta) = \beta_1+\beta_2x^2$, which is one of the choices in \cite{leisen2017bayesian}.

We compare the performance of both models by means of an estimate to the integrated $L_1$ distance, denoted $\widehat{IL}_1$,  between the true conditional  Kendall's tau function, denoted $\tau^{true}(x)$, and its conditional estimate, denoted $\widehat{\tau}(x)$, given by
\begin{align}\label{integratedL1}
\widehat{IL}_1&= \frac{1}{L}\sum_{l=1}^L\left\vert \widehat{\tau}(x_l) - \tau^{true}(x_l) \right\vert,
\end{align}
where $\widehat{\tau}$ is given by the pointwise posterior predictive median and  $\{x_l\}_{l=1}^L$ denotes an equally spaced grid of values on $(0,1)$.

Under both simulated scenarios the specification of the DDPMC model is completed by considering a truncation level of $N$ for the random measure. Since only one  covariate is simulated, $p=2$, and the hyperparameters were set to  $\bmu^v=\bmu^{\rho}=(0,0)^t$ and $\Sigma^v=\Sigma^{\rho}=2.25\bI_2$, where $\bI_p$ denotes the $(p\times p)$ identity matrix.  Under this specification, the prior distribution induced for the stick-breaking variables assigns positive probability to the $(0, 1)$ interval and the prior distribution induced for the correlation parameter is concentrated on the $(-0.70, 0.99)$ interval, for varying values of the predictor.  Note that this prior distribution for the correlation parameter is adequate for both simulation scenarios.   Following a similar model specification, for the LDVR  model, we consider  $\bmu=(0,0)^t$ and $\Sigma=2.25\bI_2$, which also results in a prior distribution for the correlation parameter that is concentrated on the $(-0.70, 0.99)$ interval,  and $\alpha=1$.  
See the Supplementary Material for more details on the selection of these hyperparameters.

The Markov chain Monte Carlo (MCMC) posterior sampling scheme for the proposed DDPMC model, sequentially updates the coefficients of the linear predictors defining $v_j$ and $\rho_j$ with a multivariate slice sampler step (\cite{neal;2003}). For a complete detail on the sampling scheme see the Supplementary Material. 
We use the recently developed BNP functionality of the \texttt{nimble} package (\cite{de2017programming}, \cite{wehrhahn2018bayesian}) from the \texttt{R} statistical software to generate posterior samples of the  LDVR  model. The sampling scheme is based on a truncation to a level $N$ of the random measure and on random walk Metropolis-Hastings and categorical samplers for the parameters in the model.

Under both models,  a single Markov chain was generated for each simulated data set. For the proposed DDPMC model a Markov chain of size 110,000 was generated and  inference is based on a reduced chain of size 10,000 obtained by saving one every 10 iterations, after a burn-in period of 10,000, for both data sets.  For the LDVR  model,  Markov chains of size 1,510,000 and 210,000 were generated for Scenarios I and II, respectively. For both scenarios, inferences are based on a reduced chain of size 10,000 obtained after a burn-in period of 10,000 and thinning every 150 and 20 iterations for Scenarios I and II, respectively. Convergence of the posterior sample was evaluated with standard convergence tests as implemented in the CODA R library \cite{plummer2006coda} and by examining trace plots.

\begin{figure}[bt]
\centering
\begin{tabular}{ccc}
&Scenario I & Scenario II \\
\rot{\hspace{1.8cm} \large{DDPMC model}} &
\subfigure[]
{
    \includegraphics[scale=0.3]{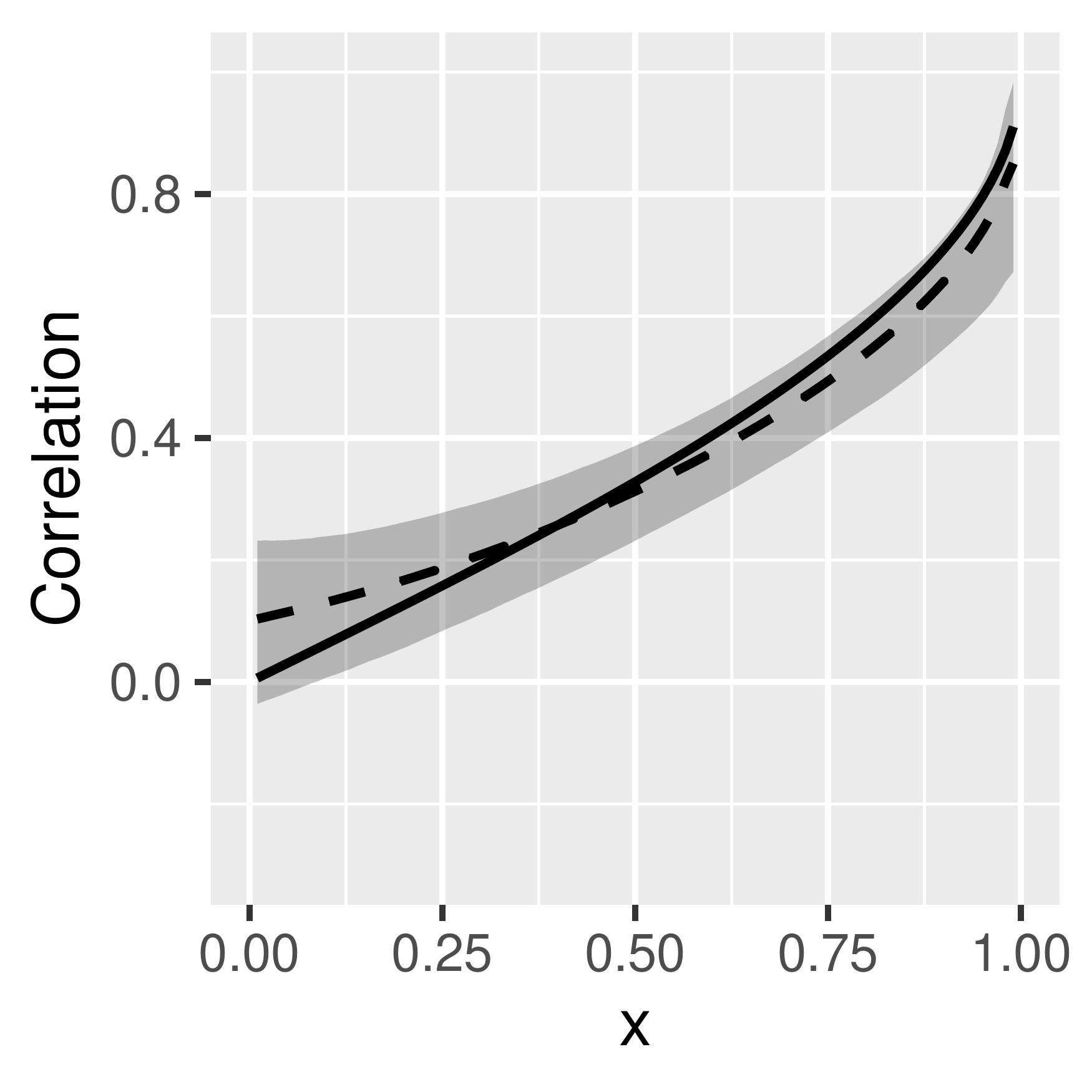}
}
&
\subfigure[]
{
    \includegraphics[scale=0.3]{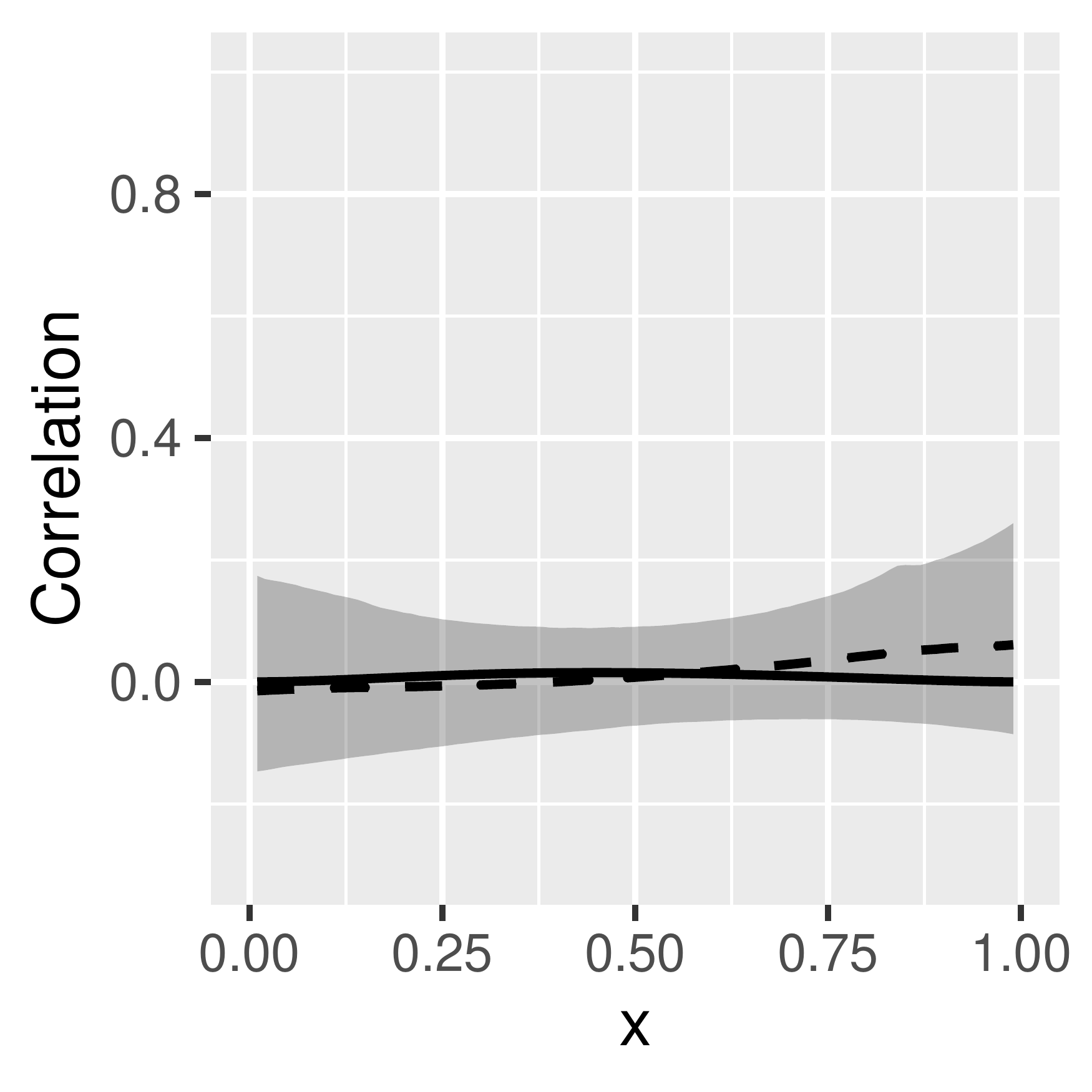}
}\\
\rot{\hspace{1.8cm} \large{LDVR  model}} &
\subfigure[]
{
    \includegraphics[scale=0.3]{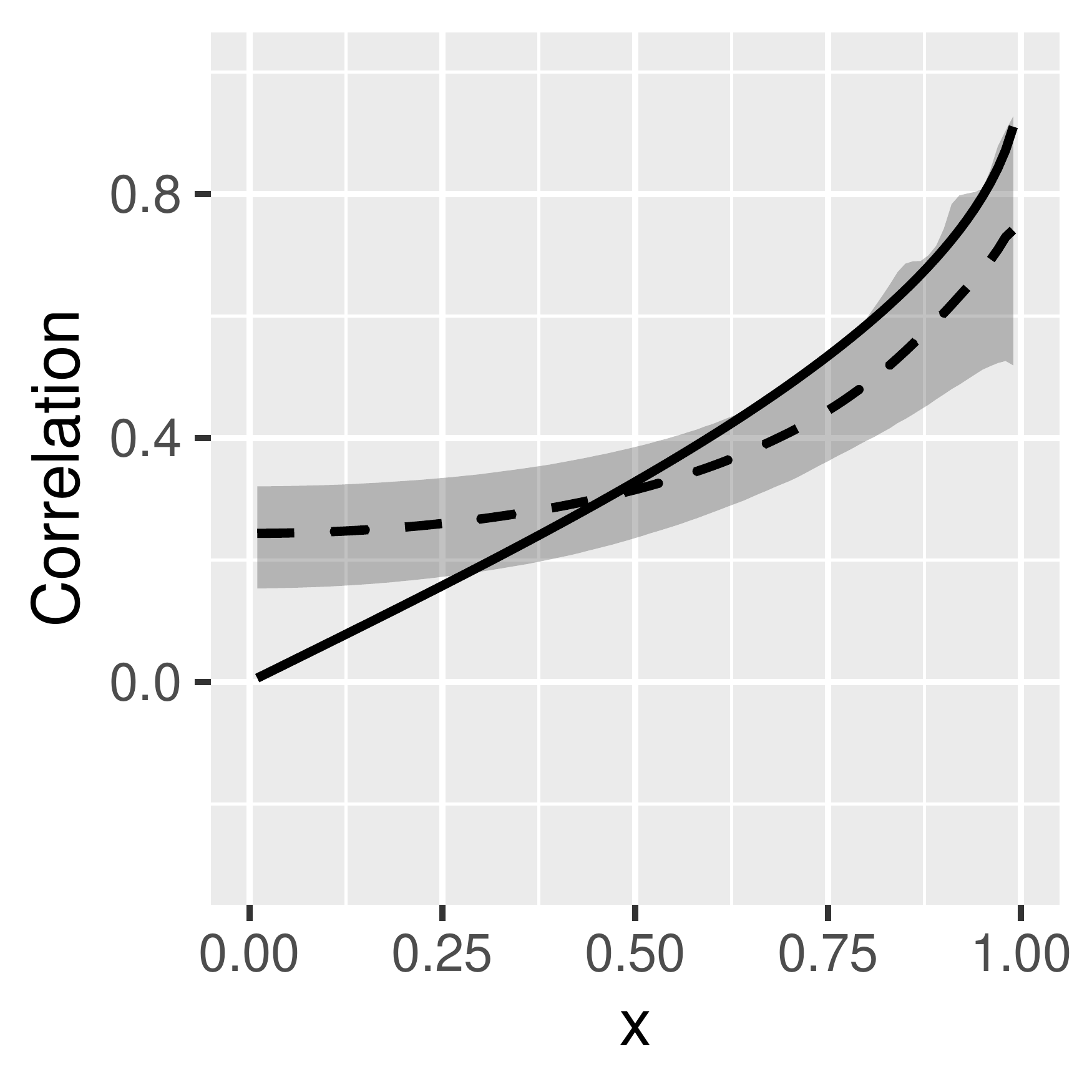}
}
&
\subfigure[]
{
    \includegraphics[scale=0.3]{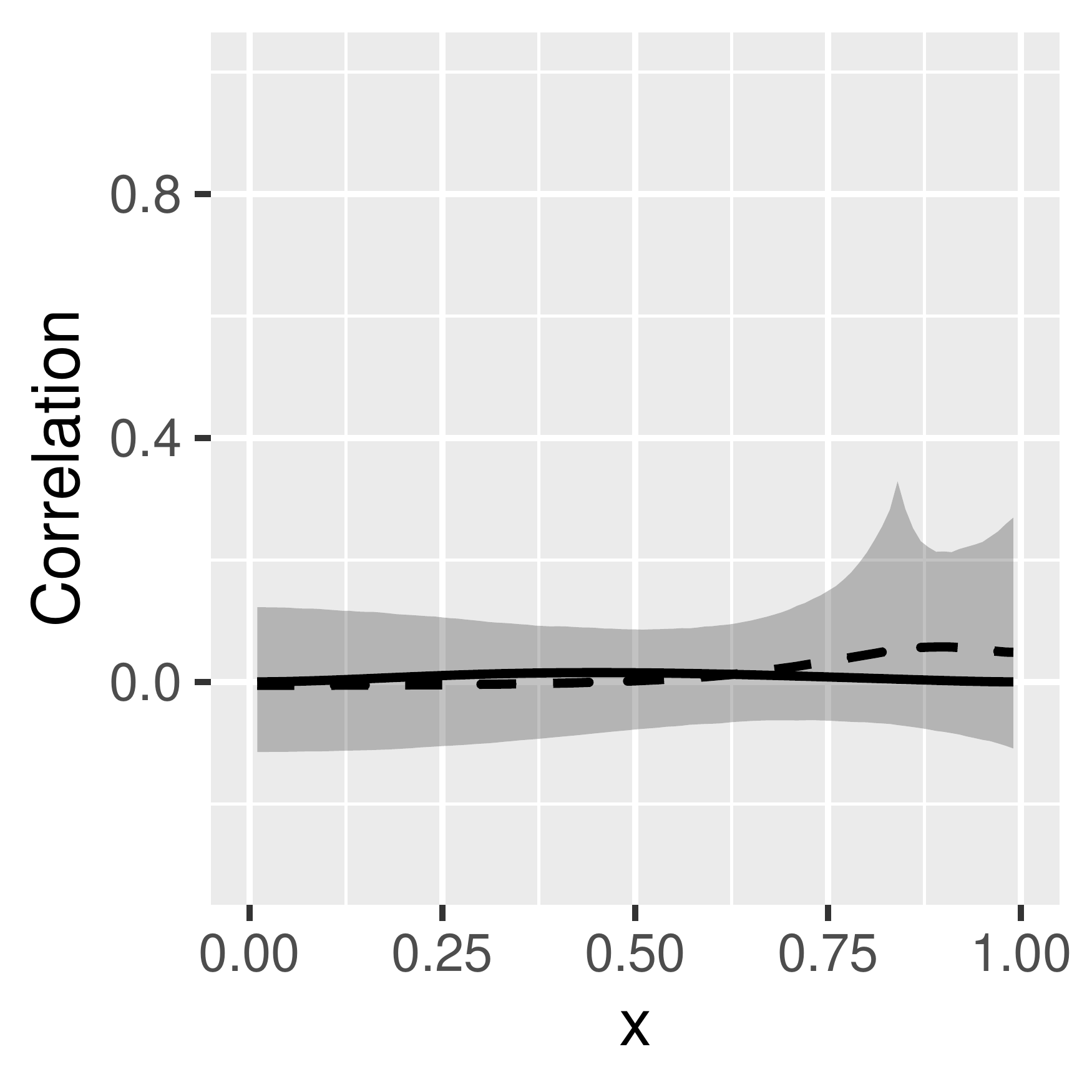}
}
\end{tabular}
\\
\caption{\label{sim:KendallsTau} }{Simulated Data: Kendall's tau estimate (dashed line) and true Kendall's tau (solid line) as a function of the predictor, $x$,  for Scenario I (first column) and Scenario II (second column). Results are displayed  for the  DDPMC model (first row) and LDVR  model (second row).   Grey areas represent a 95\% credible band.}
\end{figure}

Figure \ref{sim:KendallsTau} displays the covariate dependent Kendall's tau  estimate (pointwise posterior predictive median), its corresponding 95\% credible bands, and the true covariate dependent Kendall's tau for Scenario I and Scenario II and under both models.
Note that under Scenario I, the proposed DDPMC model is  able to  recover the  dynamic  association structure between the random variables for every value of the predictor, while the  LDVR  model is not able to recover the true association structure over all possible values of the covariate. The  LDVR  model provides good estimates for values greater than 0.3 of the covariate, but struggles for   smaller values of the covariate. 
Under the proposed DDPMC model, the true values of Kendall's tau lie within the 95\% credible band for every value of the predictor, which is not the case for  the  LDVR  model. 
  Under Scenario II, Kendall's tau is estimated as an almost constant function of the predictor, with very low values.  The true Kendall's tau lies within the 95\% credible band for every value of the predictor, under both modeling approaches. 
See the Supplementary Material  for additional results displaying the contour and surface plots of the copula density estimates and true copula density for selected values of the predictor. 

In order to quantify the differences of the results using both methodologies, we  compute estimates to the integrate $L_1$ distances, given by Equation (\ref{integratedL1}). For simulation Scenario I these values are $0.0387$ and $0.0934$, under the proposed DDPMC model and the LDVR  model, respectively.   Correspondingly, for Scenario II these values are $0.0210$ and $0.0214$. These results are consistent with the previous discussion.


\section{Application}
\label{sec:appl}

Let $Y_{i1}$ and $Y_{i2}$ be  measurements of  cardiovascular   and hyperglycemic risk markers of a normoglycemic patient $i$ from the MCDS, respectively.  Here, measurements of CIMT represent the cardiovascular risk marker, and measurements of HbA1c and PG2-H represent the  hyperglycemic markers.  In order to determine which hyperglycemic marker shows a stronger relationship with cardiovascular risk,  we model the dependence structure between each one and the CIMT. Additionally, we consider that such dependence structure is affected  by the triglycerides level, BMI, age, and   sex of  patient $i$, encoded in a predictor vector $\bx_i$.

 For both real data sets the specification of the model is completed by considering a truncation level of $N$ for the random measure. Since the predictor age was discretized into four categories the real data sets contain $p=7$ predictors. Here, the mean hyperparameters were set to  $\bmu^v=\bmu^{\rho}=\boldsymbol{0}^t$ and the covariances $\Sigma^v$ and $\Sigma^{\rho}$ were defined by block diagonal matrices of the form 
 \begin{equation*}
\Sigma^v=\begin{pmatrix}
    c_1^v(\tilde{\mathbb{X}}^t\tilde{\mathbb{X}})^{-1}       &\boldsymbol{0} \\
   \boldsymbol{0}       &c_2^v(\overline{\mathbb{X}}^t\overline{\mathbb{X}})^{-1} \\
    \end{pmatrix} \quad \mbox{and}  \quad
    \Sigma^{\rho}=\begin{pmatrix}
    c_1^{\rho}(\tilde{\mathbb{X}}^t\tilde{\mathbb{X}})^{-1}       &\boldsymbol{0} \\
   \boldsymbol{0}       &c_2^{\rho}(\overline{\mathbb{X}}^t\overline{\mathbb{X}})^{-1} \\
\end{pmatrix},
\end{equation*}
  where $\tilde{\mathbb{X}}$ and $\overline{\mathbb{X}}$ denote the design matrices based on the continuous  and discrete  covariates, respectively (\cite{zellner1986assessing}). The scaling constants $c_j^v$ and $c_j^{\rho}$, $j=1,2$,  control the linear predictor and were selected  so that the stick-breaking variables range between 0.02 and 0.99 and the correlations range between -0.70 and 1.00, for every value of the predictor, respectively. For more details on the selection of these hyperparameters, see the Supplementary Material.

As in the simulation scenario, the MCMC posterior sampling scheme sequentially updates the coefficients of the linear predictors defining $v_j$ and $\rho_j$ with a multivariate slice sampler step.  A single Markov chain of size 300,000 was generated for each real data set and inferences are based on a reduced chain of size 5,000 obtained after discarding the first 200,000 iterations and saving one every 20 iterations. Convergence of the posterior sample was evaluated with standard convergence tests as implemented in the CODA R library and by examining trace plots. 

\begin{figure}
\centering
\scalebox{0.55}{
\begin{tabular}{ccccc}
&\large{$age < 55$} & \large{$55 \leq age < 60$} & \large{$60 \leq age < 65$}  & \large{$age \geq 65$} \\
\rot{\hspace{1.8cm} \large{Male}} &
{
    \includegraphics[scale=0.3]{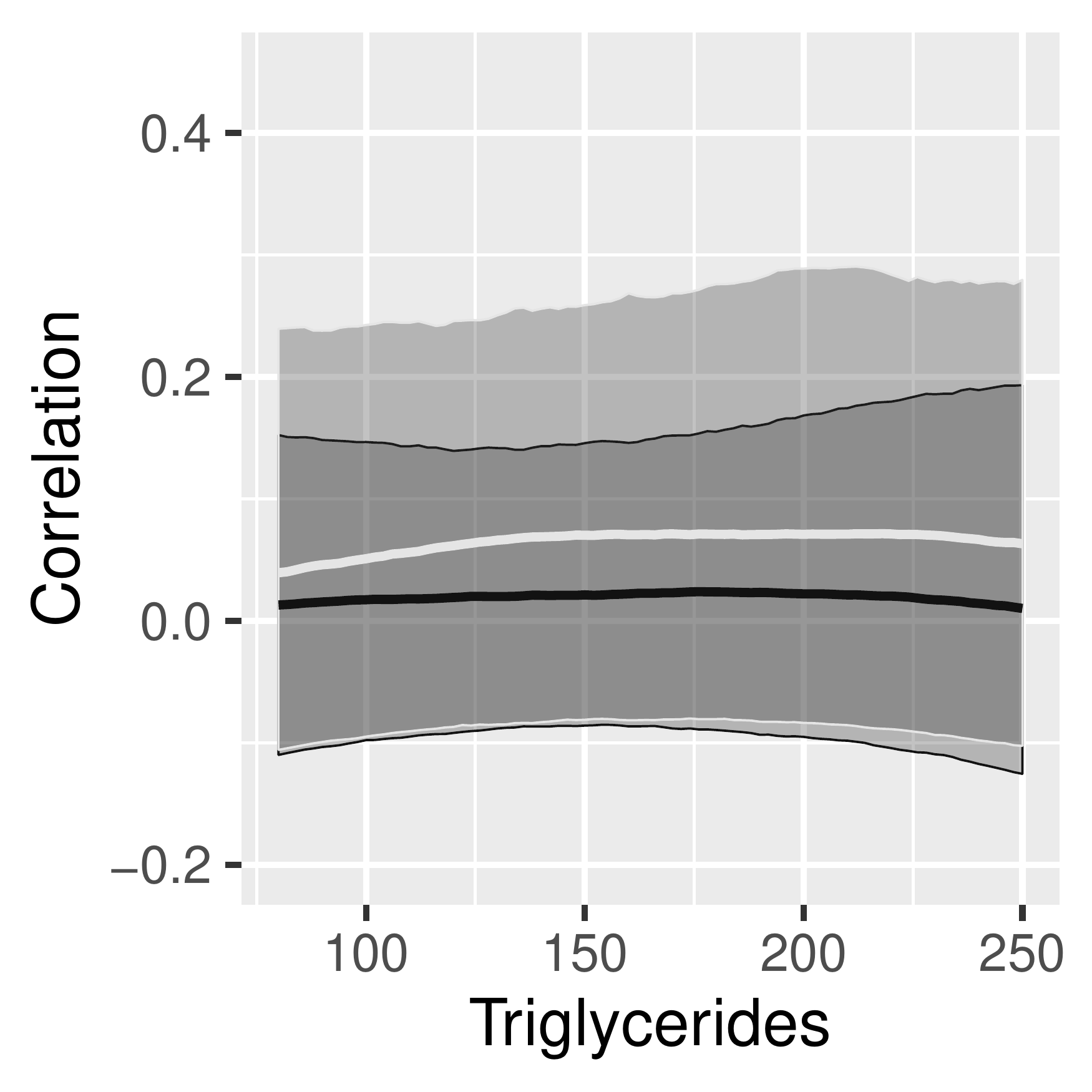}
}
&
{
    \includegraphics[scale=0.3]{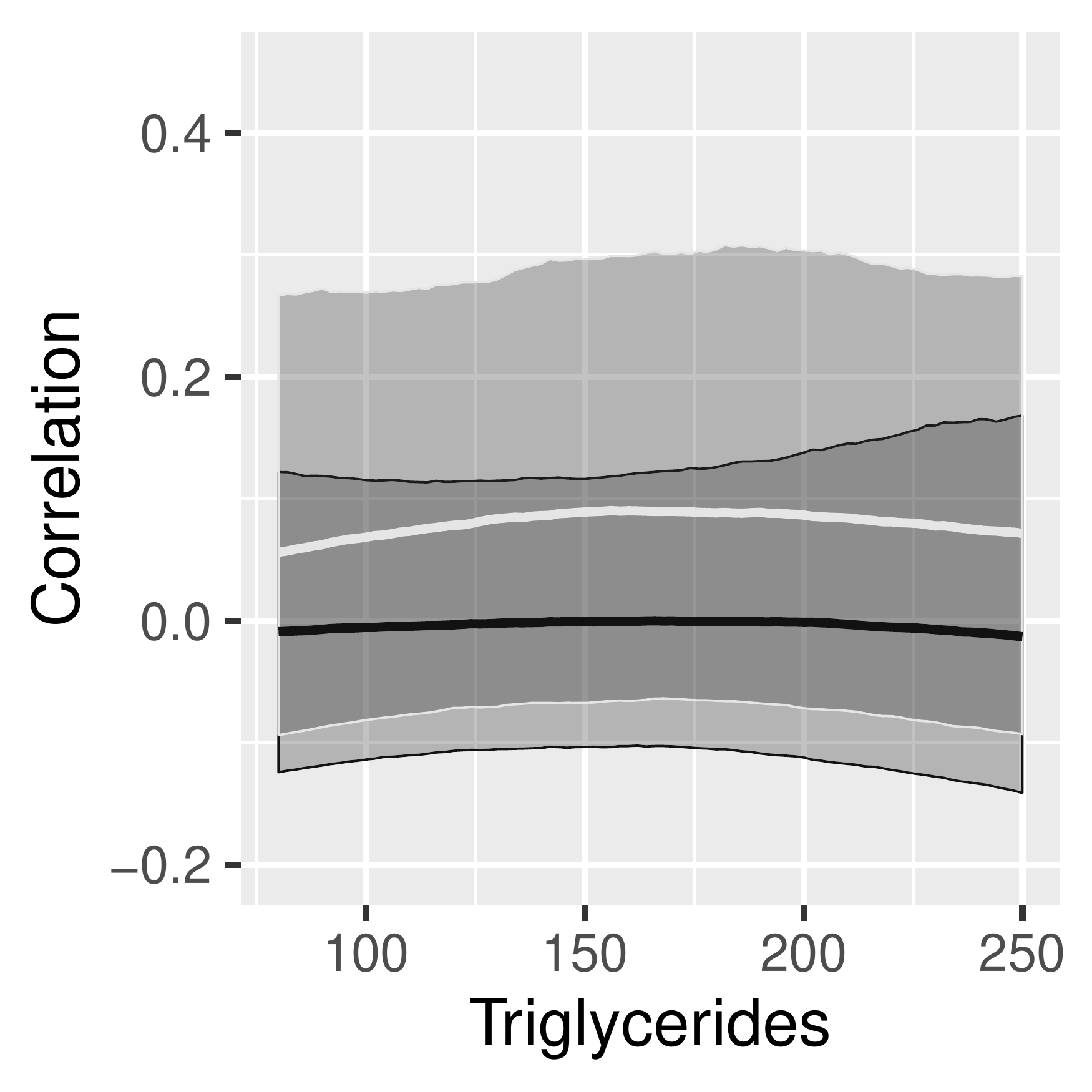}
}
&
{
    \includegraphics[scale=0.3]{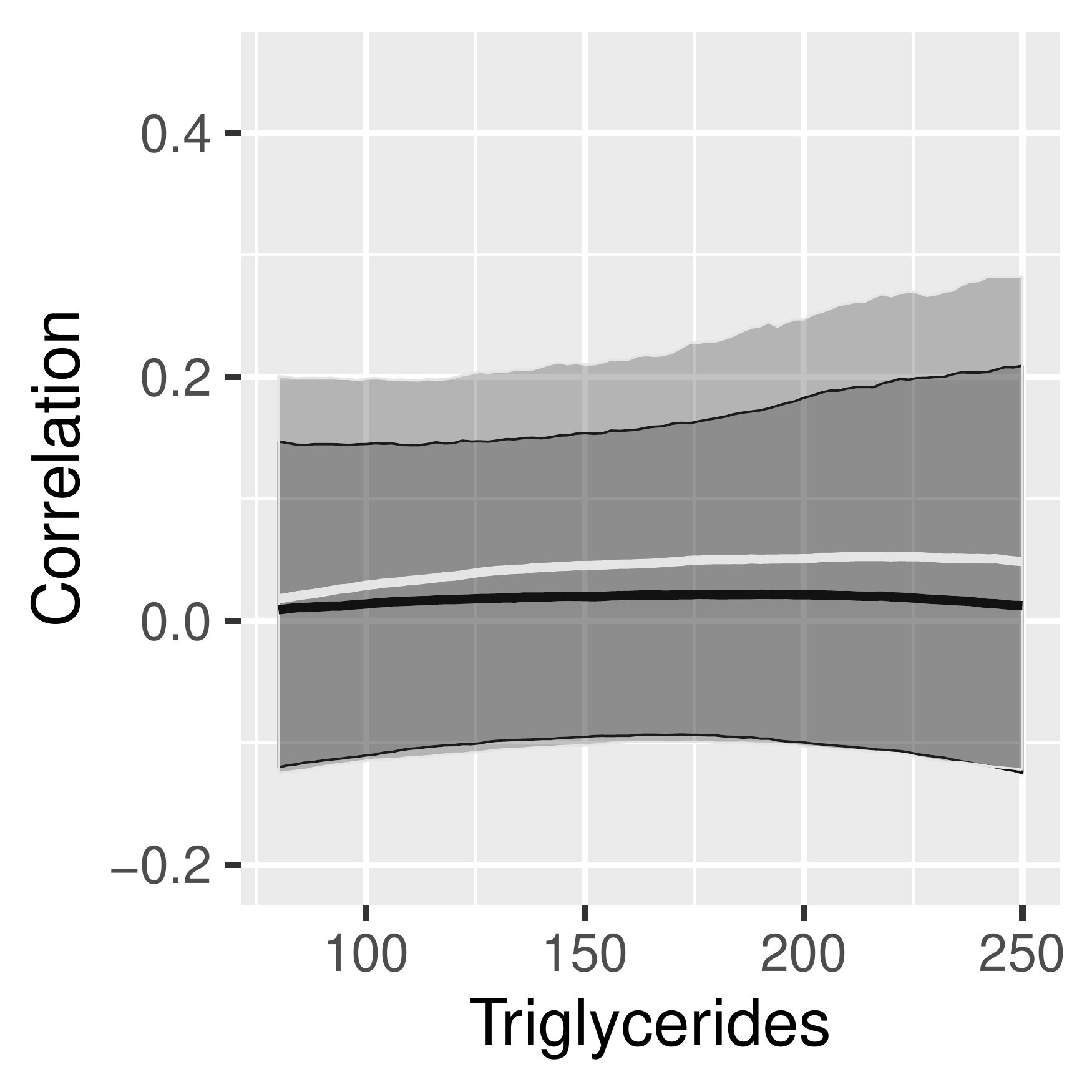}
}
&
{
    \includegraphics[scale=0.3]{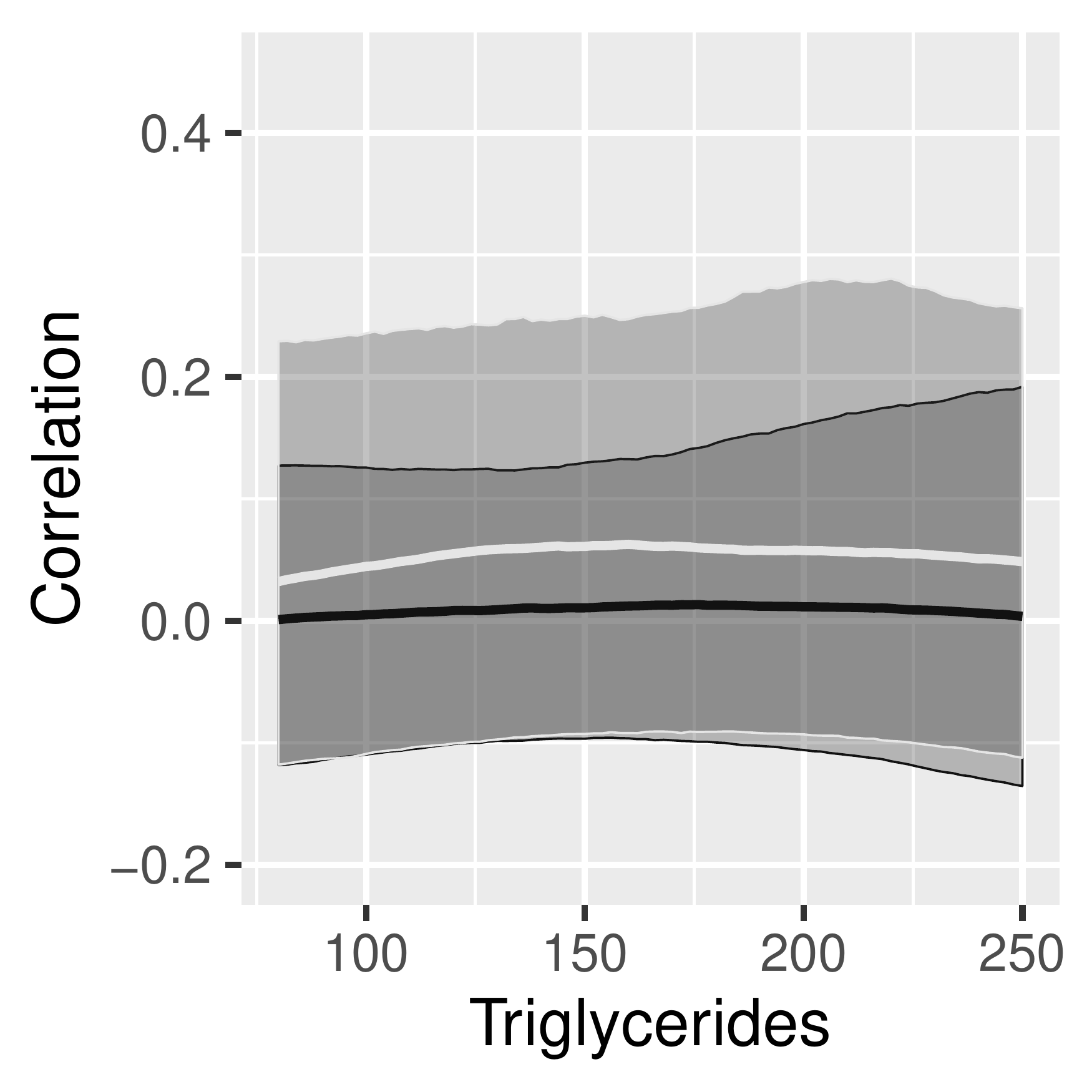}
} 
\\
\rot{\hspace{1.8cm}  \large{Female}}
&
{
    \includegraphics[scale=0.3]{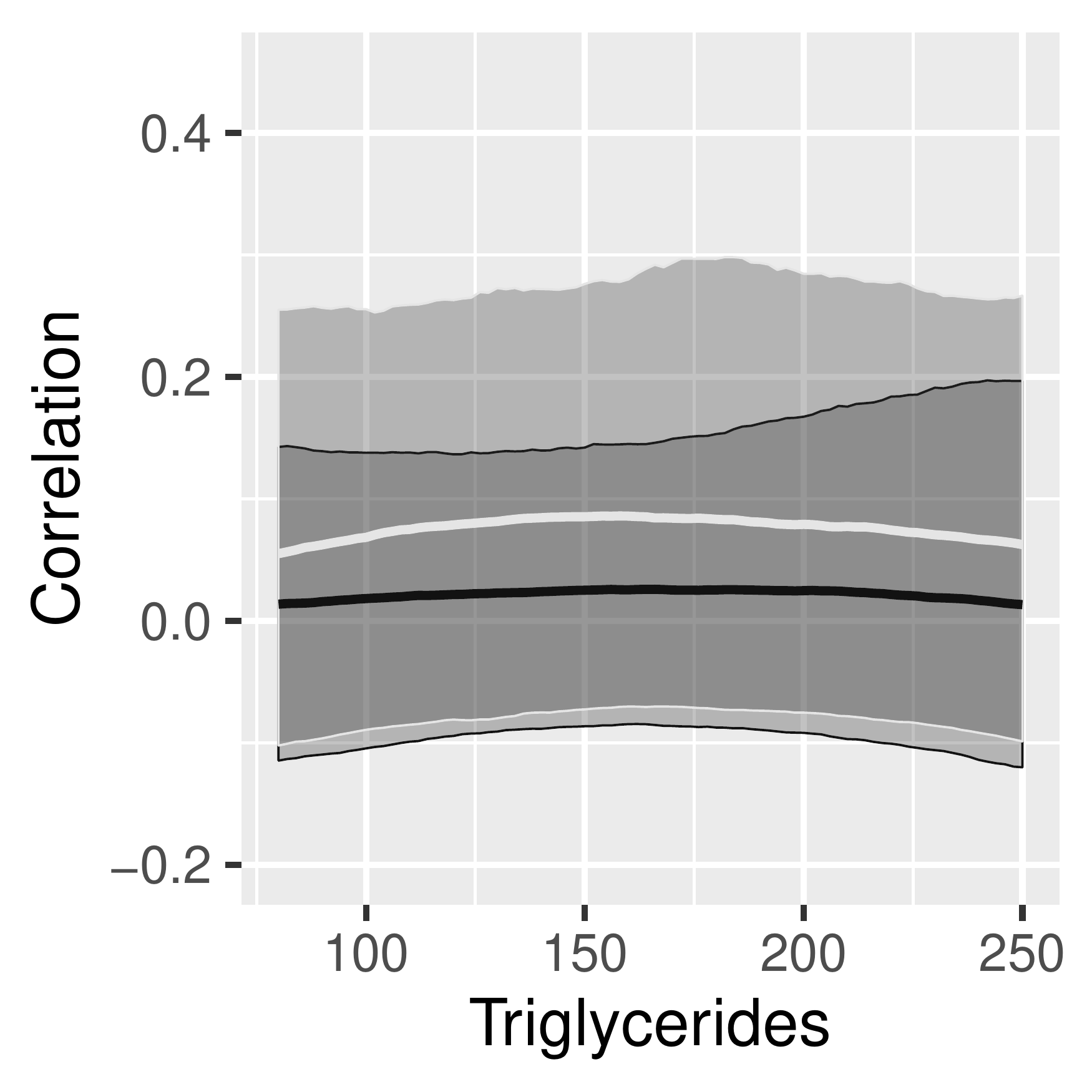}
} 
&
{
    \includegraphics[scale=0.3]{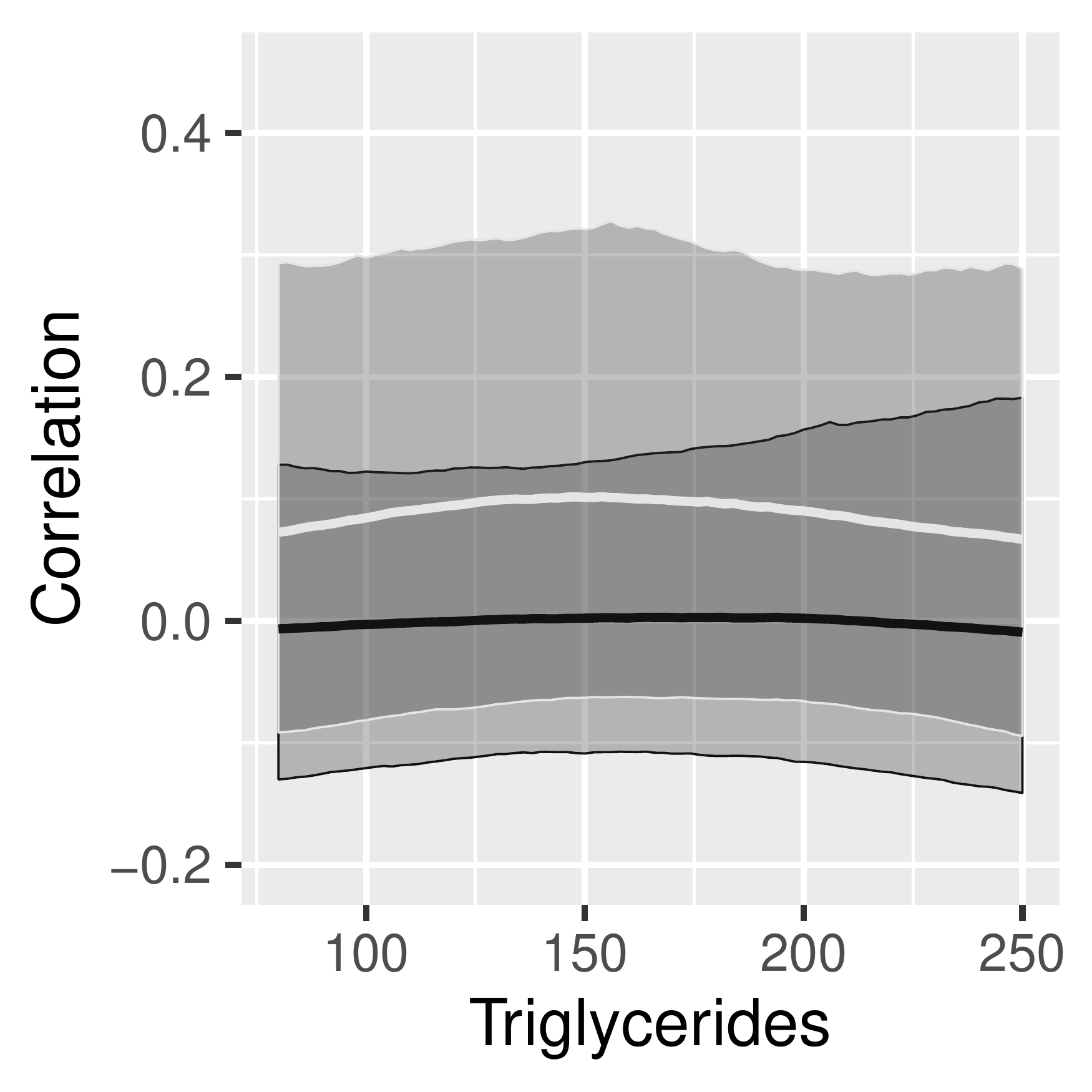}
}
&
{
    \includegraphics[scale=0.3]{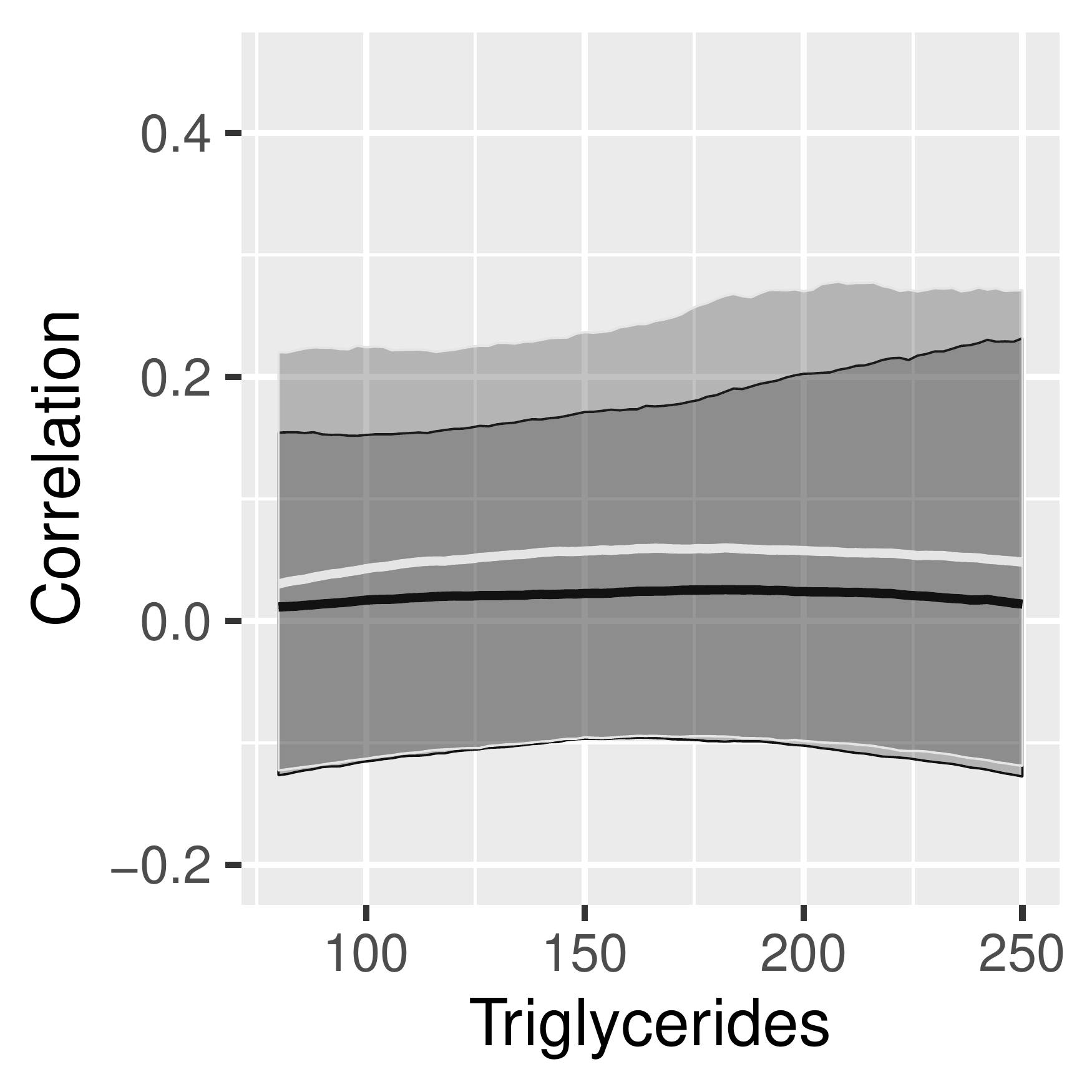}
}
&
{
    \includegraphics[scale=0.3]{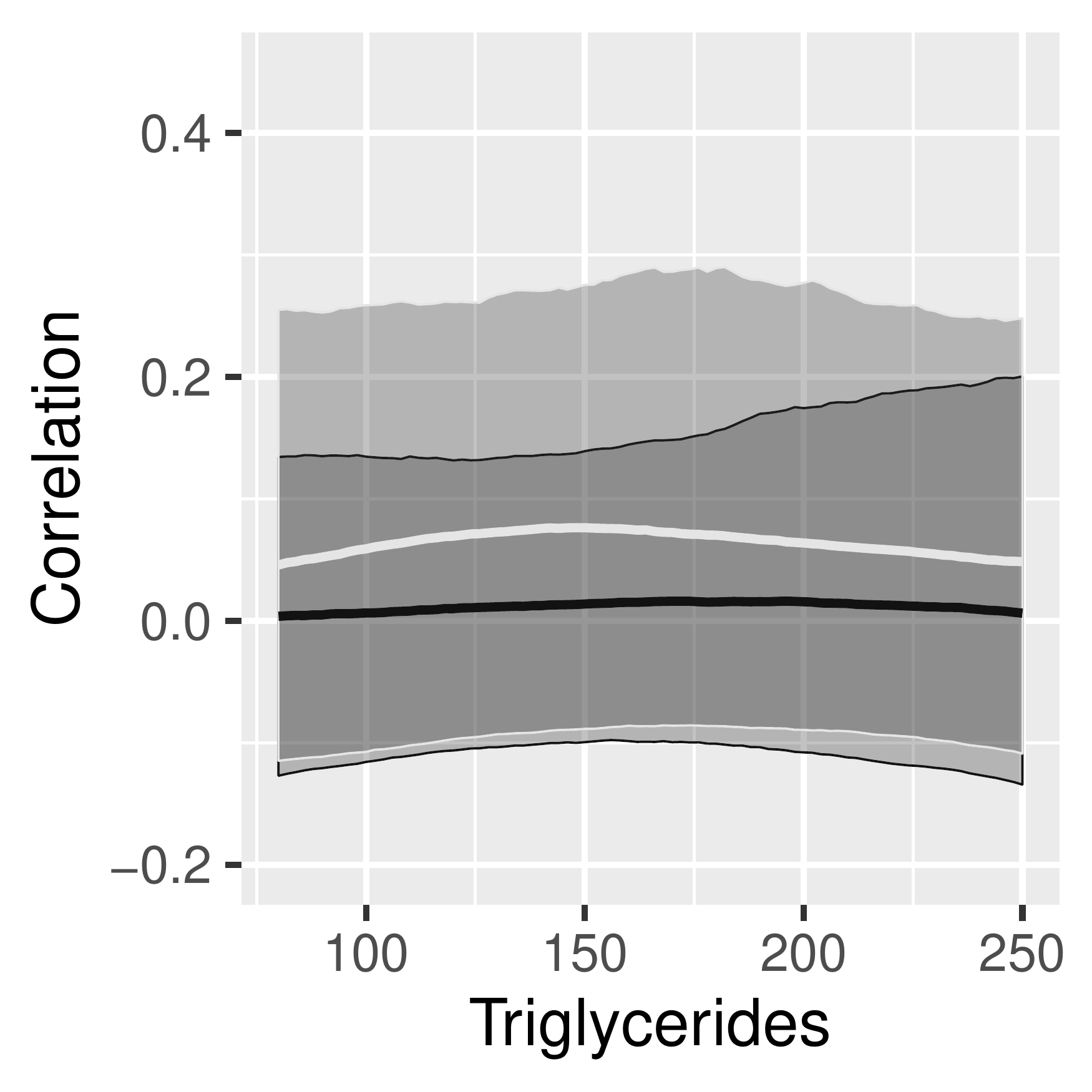}
}
\end{tabular}
}
\caption{\label{app:curveKendallsTauE1E2BMI22}\small{Mexico City Diabetes Study Data: section plot of conditional Kendall's tau  between CIMT and  PG2-H (dark grey)  and between CIMT and  Hb1Ac (light grey) when BMI  equals 22. Panels display the estimate of  Kendall's tau for varying values of triglycerides, sex, and age category for normoglycemic participants: the continuous line is the estimate (dark grey for Kendall's tau  between CIMT and  PG2-H and light grey for Kendall's tau  between CIMT and  Hb1Ac) while the grey region is a 90\% credible band. The y-axis shows values for Kendall's tau while x-axis shows values of triglycerides.}}
\end{figure}

\begin{figure}
\centering
\scalebox{0.55}{
\begin{tabular}{ccccc}
&\large{$age < 55$} & \large{$55 \leq age < 60$} & \large{$60 \leq age < 65$}  & \large{$age \geq 65$} \\
\rot{\hspace{1.8cm} \large{Male}} &
{
    \includegraphics[scale=0.3]{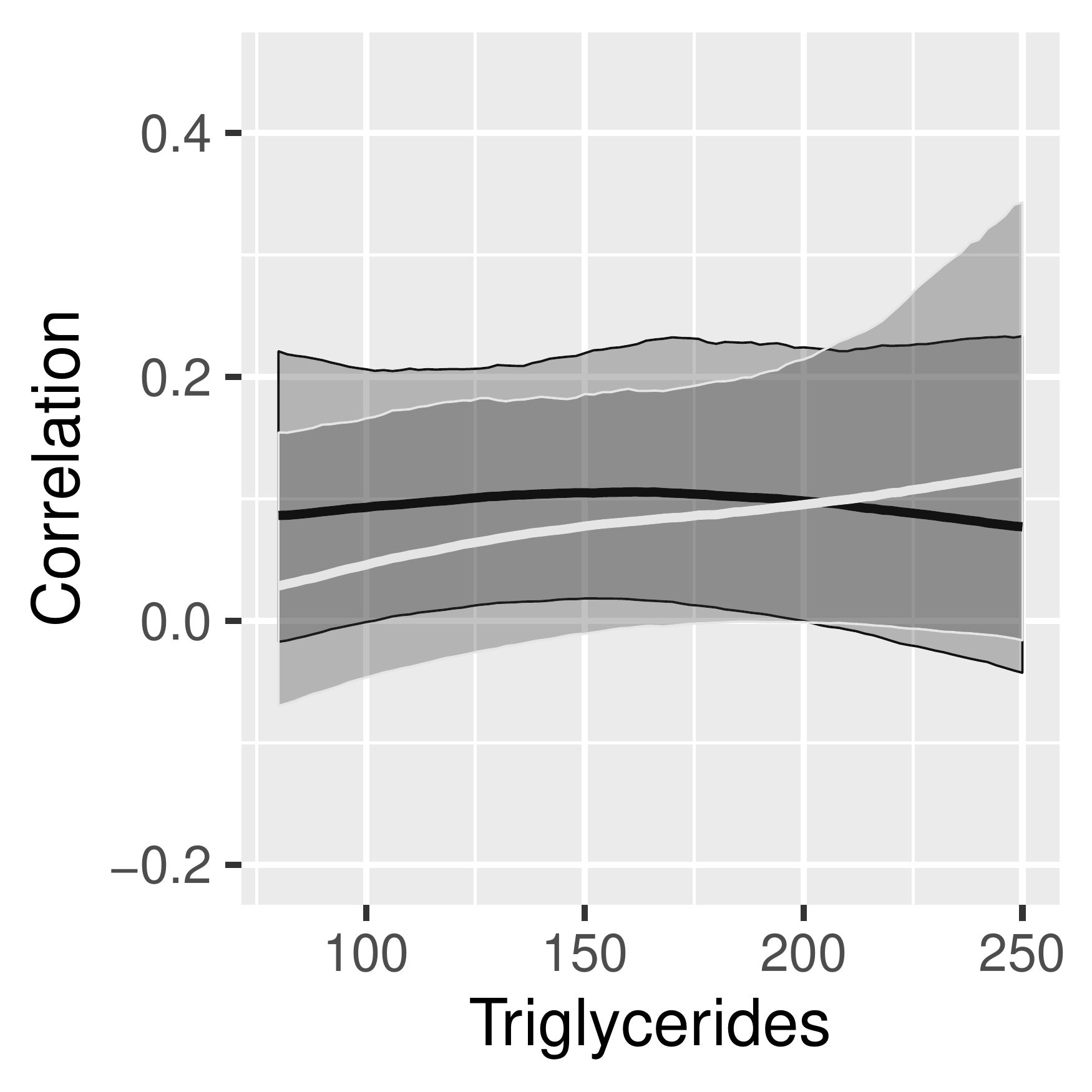}
}
&
{
    \includegraphics[scale=0.3]{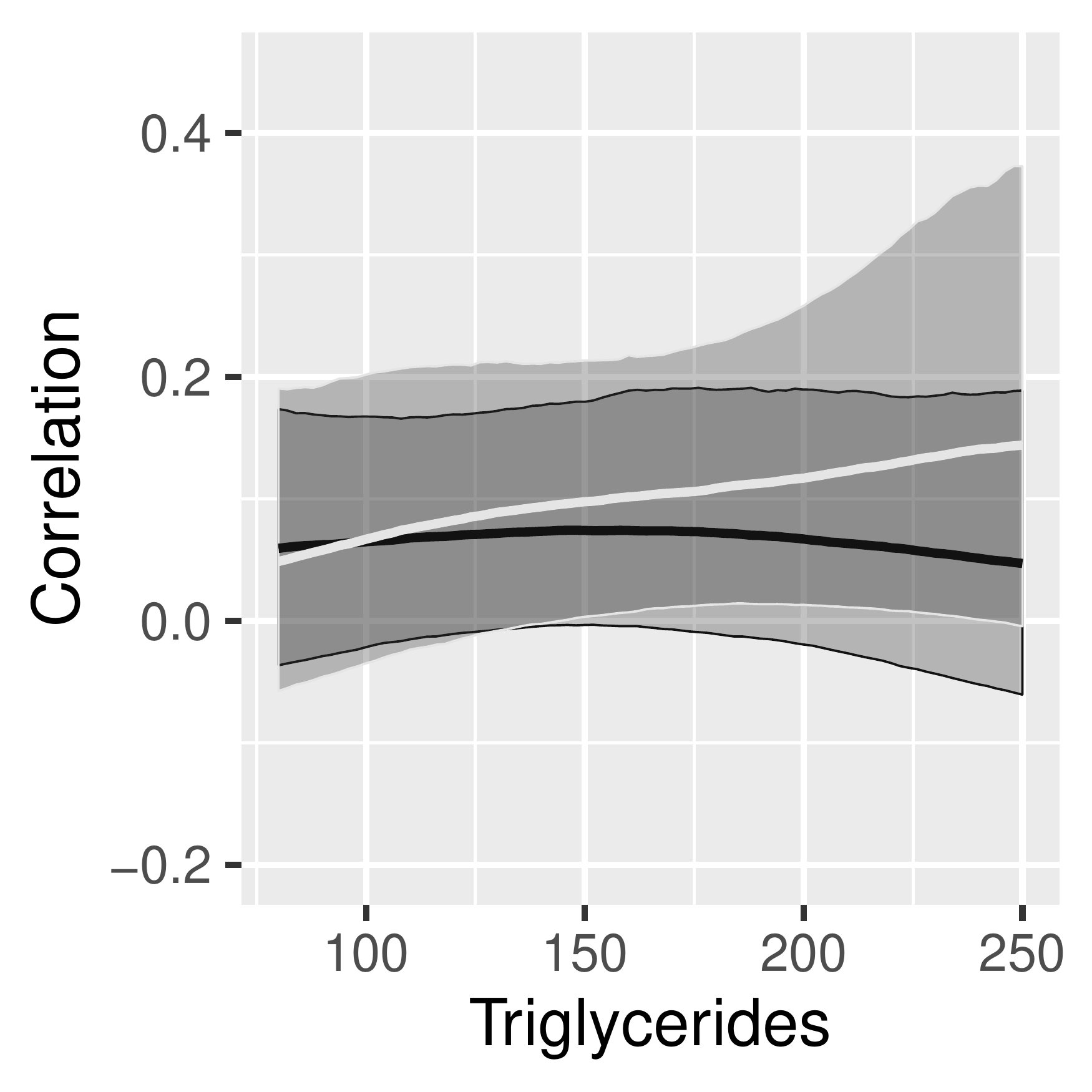}
}
&
{
    \includegraphics[scale=0.3]{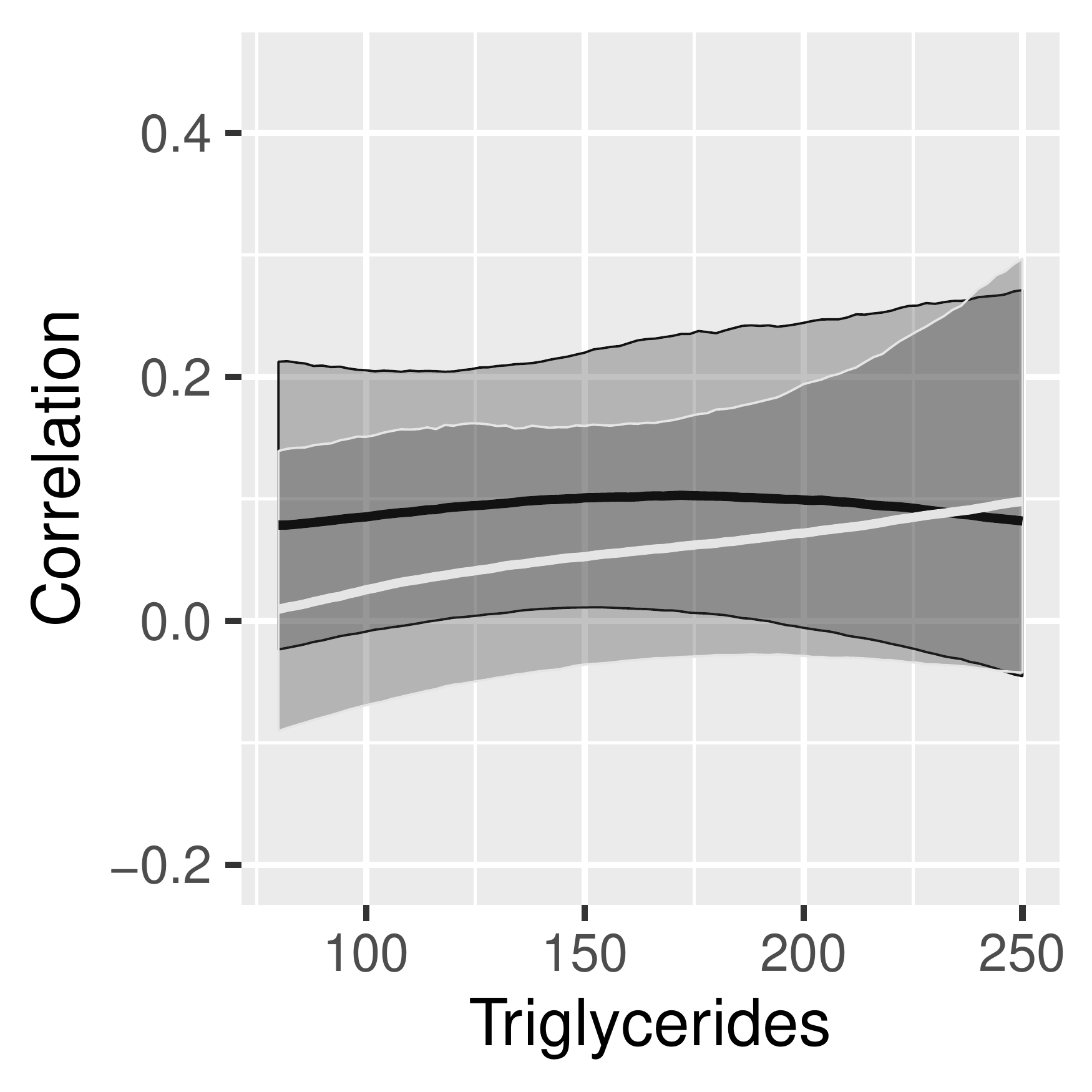}
}
&
{
    \includegraphics[scale=0.3]{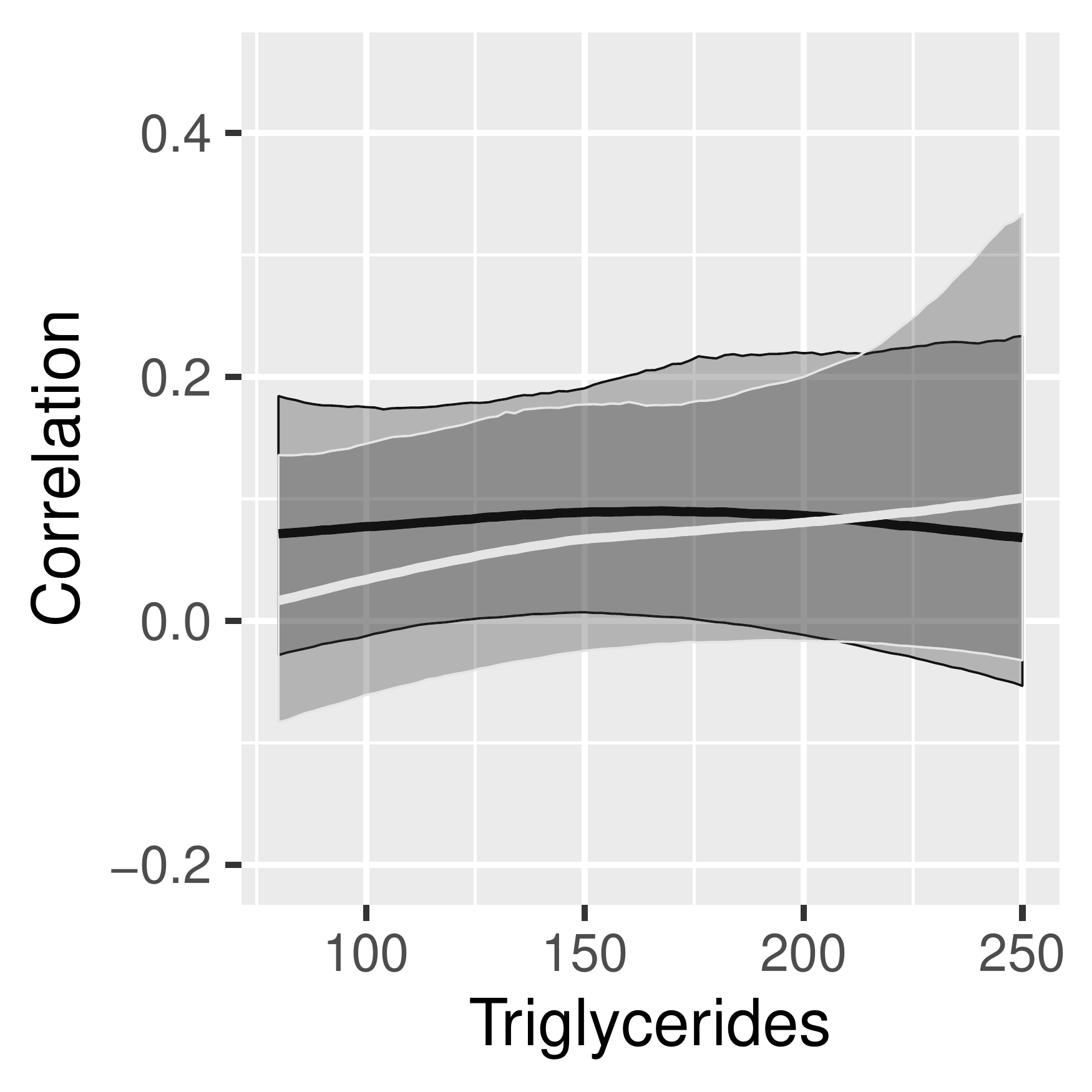}
} 
\\
\rot{\hspace{1.8cm}  \large{Female}}
&
{
    \includegraphics[scale=0.3]{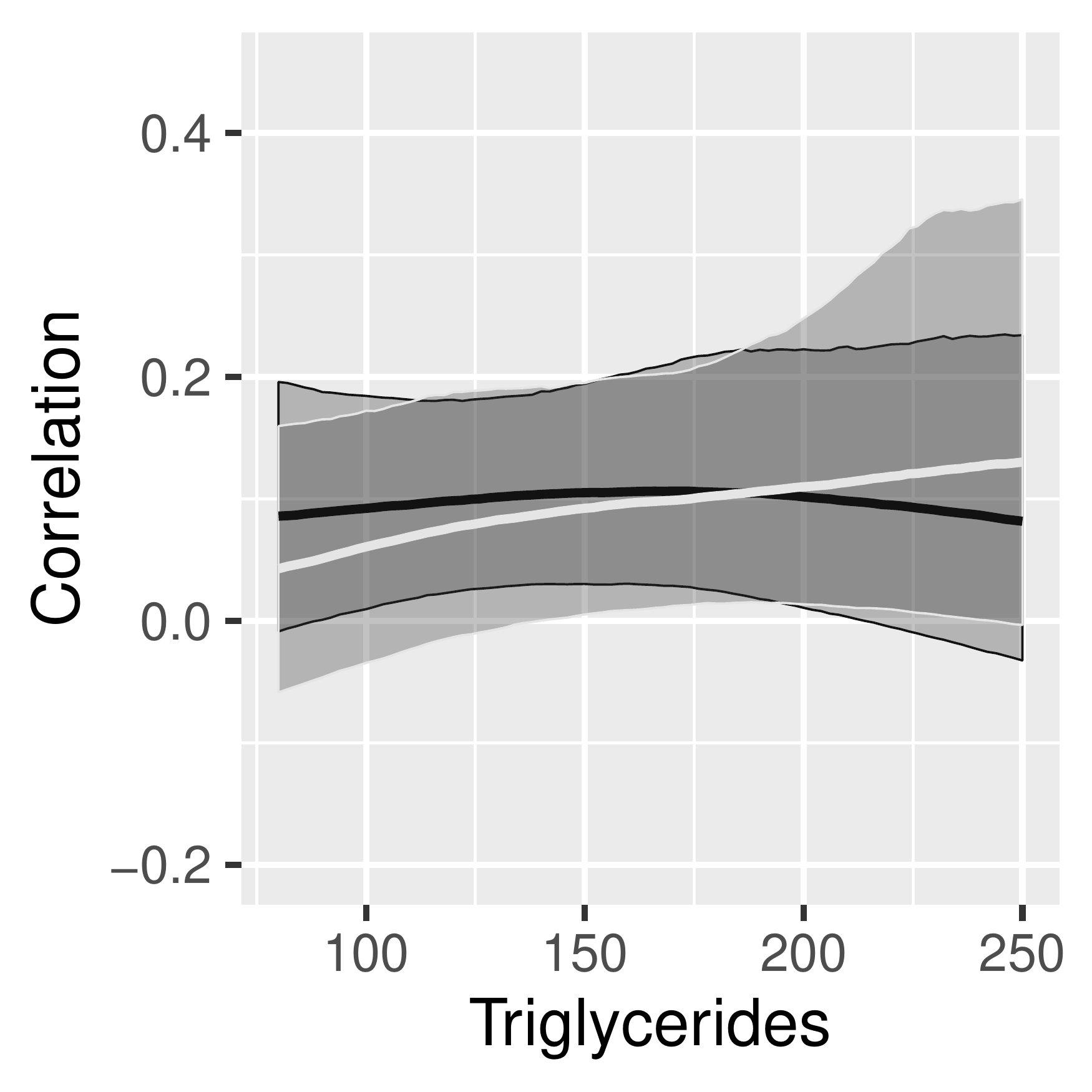}
} 
&
{
    \includegraphics[scale=0.3]{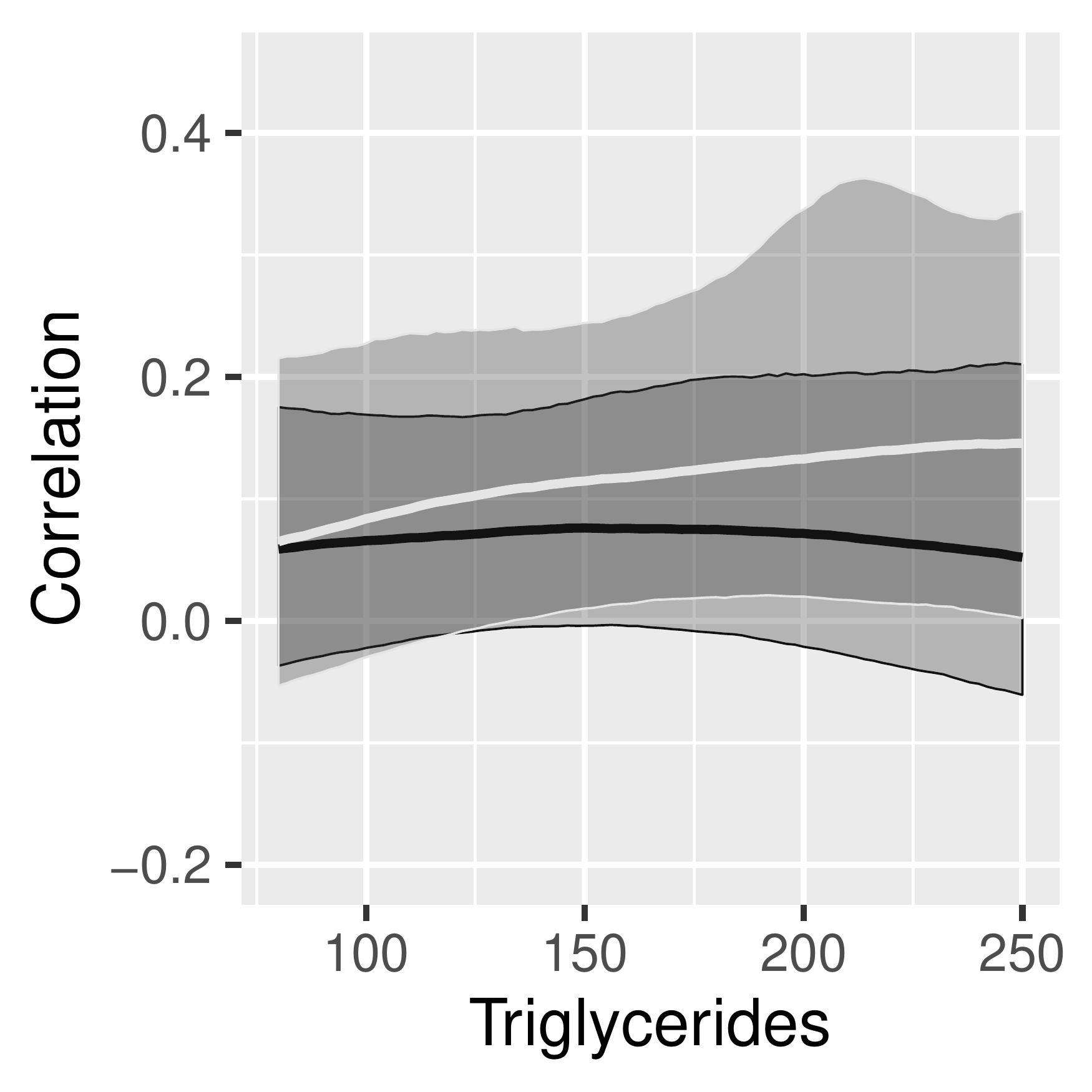}
}
&
{
    \includegraphics[scale=0.3]{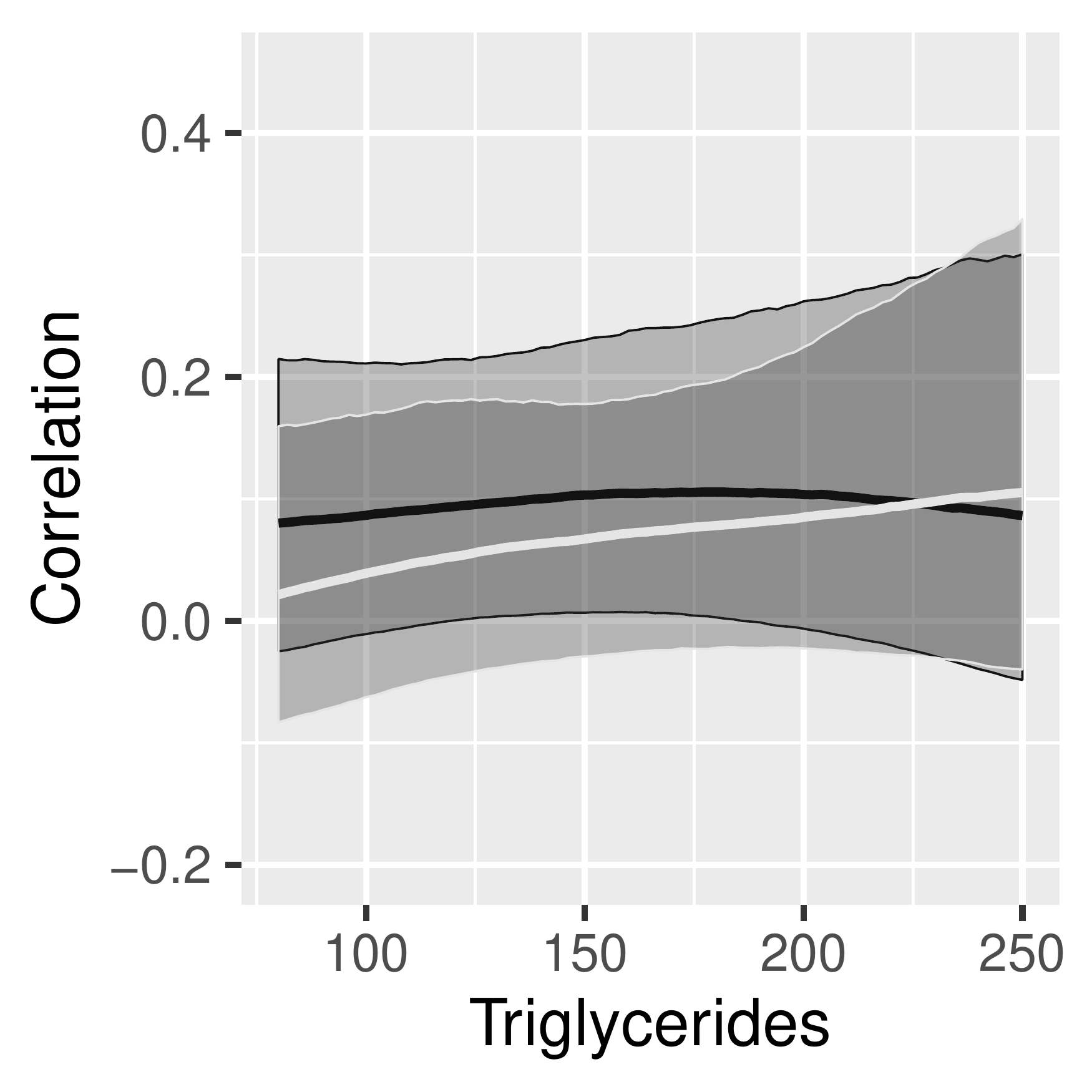}
}
&
{
    \includegraphics[scale=0.3]{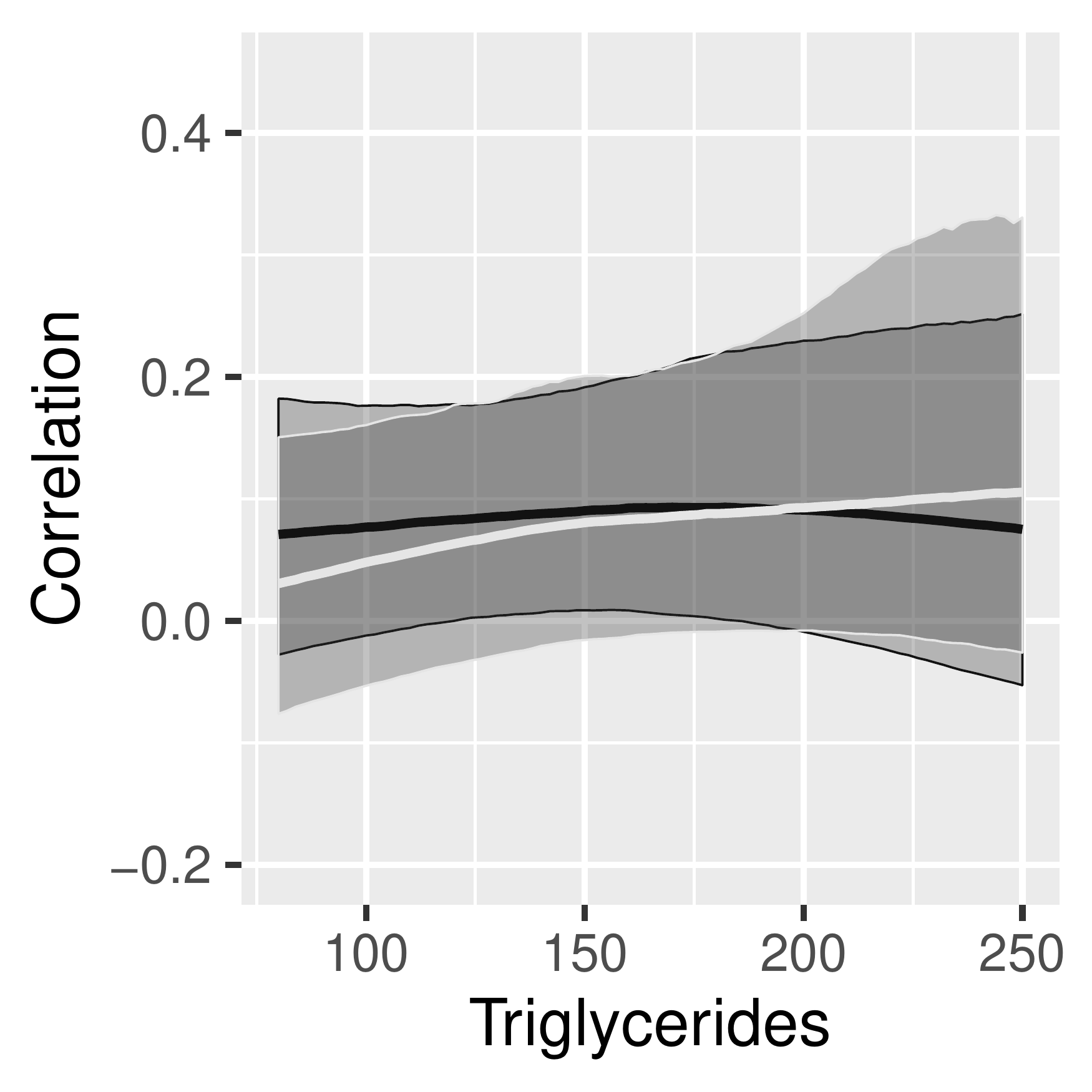}
}
\end{tabular}
}
\caption{\label{app:curveKendallsTauE1E2BMI27}\small{Mexico City Diabetes Study Data: section plot of conditional Kendall's tau  between CIMT and  PG2-H (dark grey)  and between CIMT and  Hb1Ac (light grey) when BMI  equals 27. Panels display the estimate of  Kendall's tau for varying values of triglycerides, sex, and age category for normoglycemic participants: the continuous line is the estimate (dark grey for Kendall's tau  between CIMT and  PG2-H and light grey for Kendall's tau  between CIMT and  Hb1Ac) while the grey region is a 90\% credible band. The y-axis shows values for Kendall's tau while x-axis shows values of triglycerides..}}
\end{figure}

\begin{figure}
\centering
\scalebox{0.55}{
\begin{tabular}{ccccc}
&\large{$age < 55$} & \large{$55 \leq age < 60$} & \large{$60 \leq age < 65$}  & \large{$age \geq 65$} \\
\rot{\hspace{1.8cm} \large{Male}} &
{
    \includegraphics[scale=0.3]{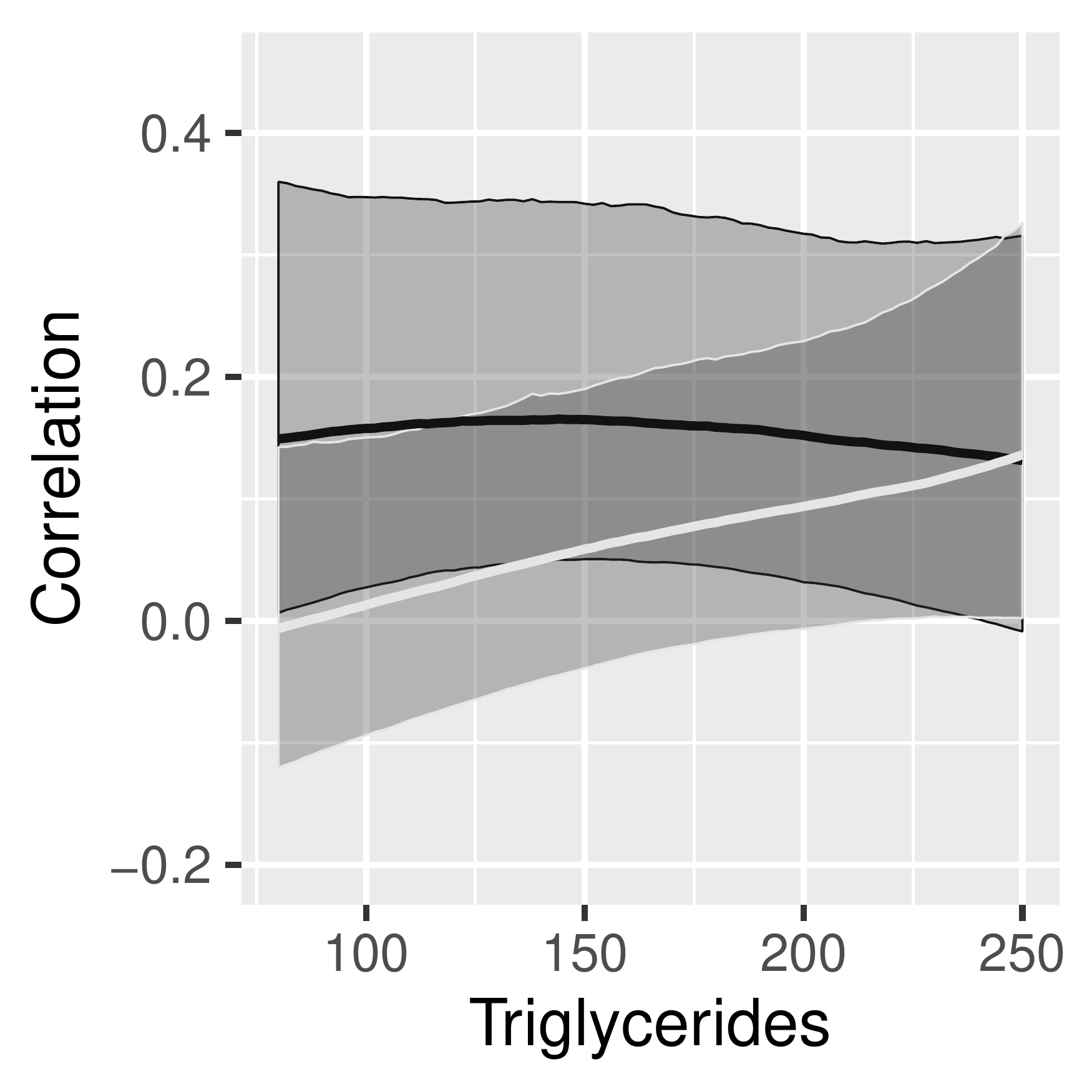}
}
&
{
    \includegraphics[scale=0.3]{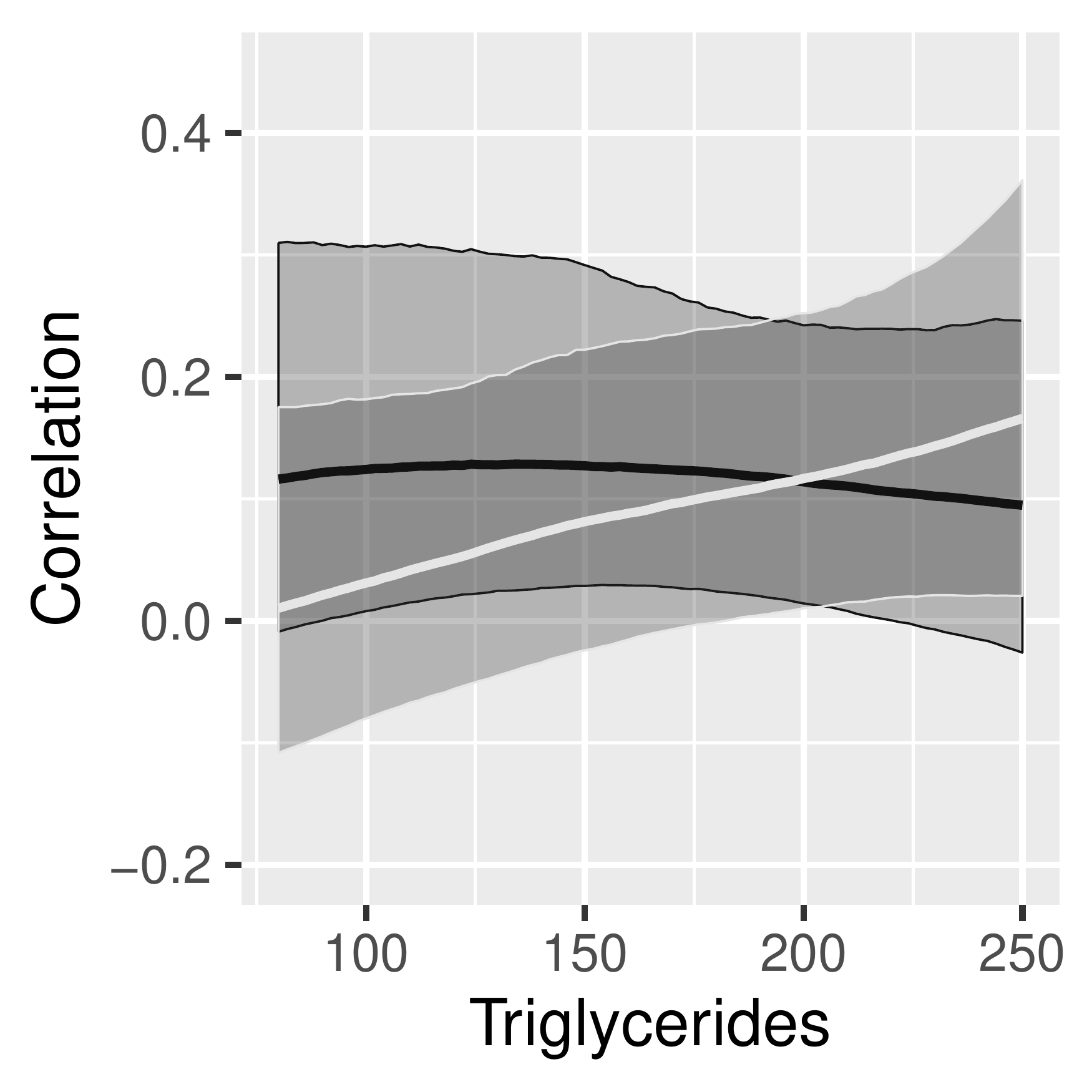}
}
&
{
    \includegraphics[scale=0.3]{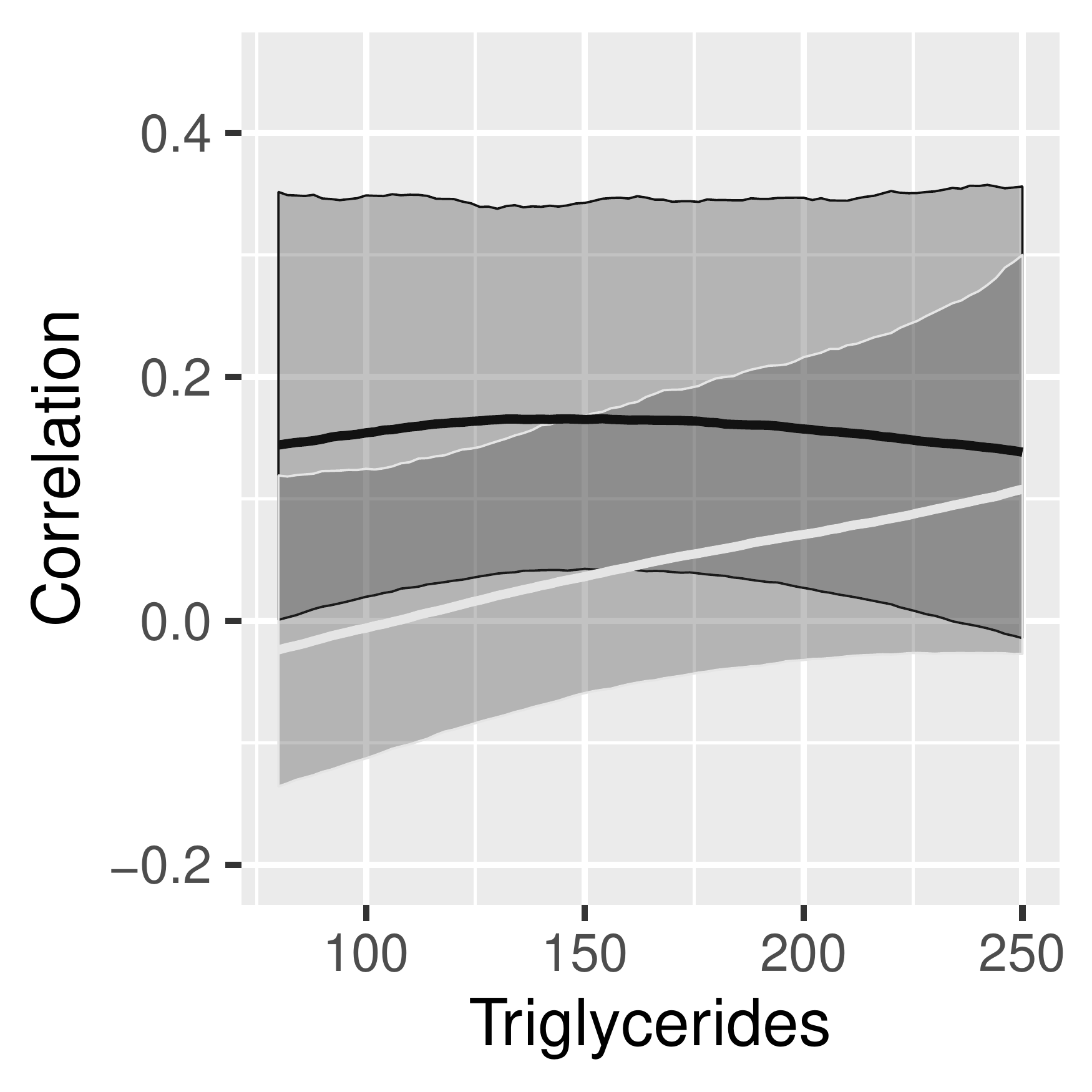}
}
&
{
    \includegraphics[scale=0.3]{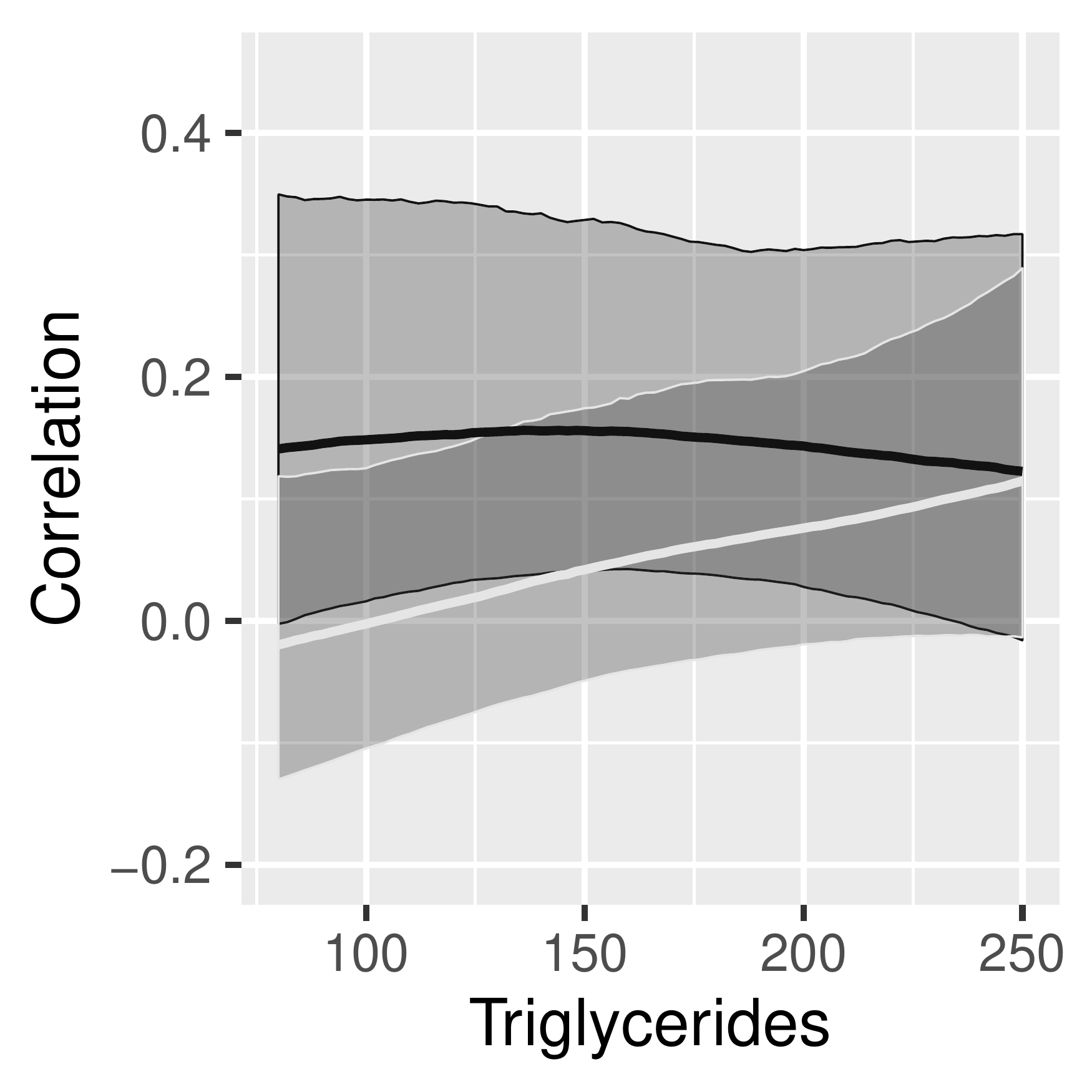}
} 
\\
\rot{\hspace{1.8cm}  \large{Female}}
&
{
    \includegraphics[scale=0.3]{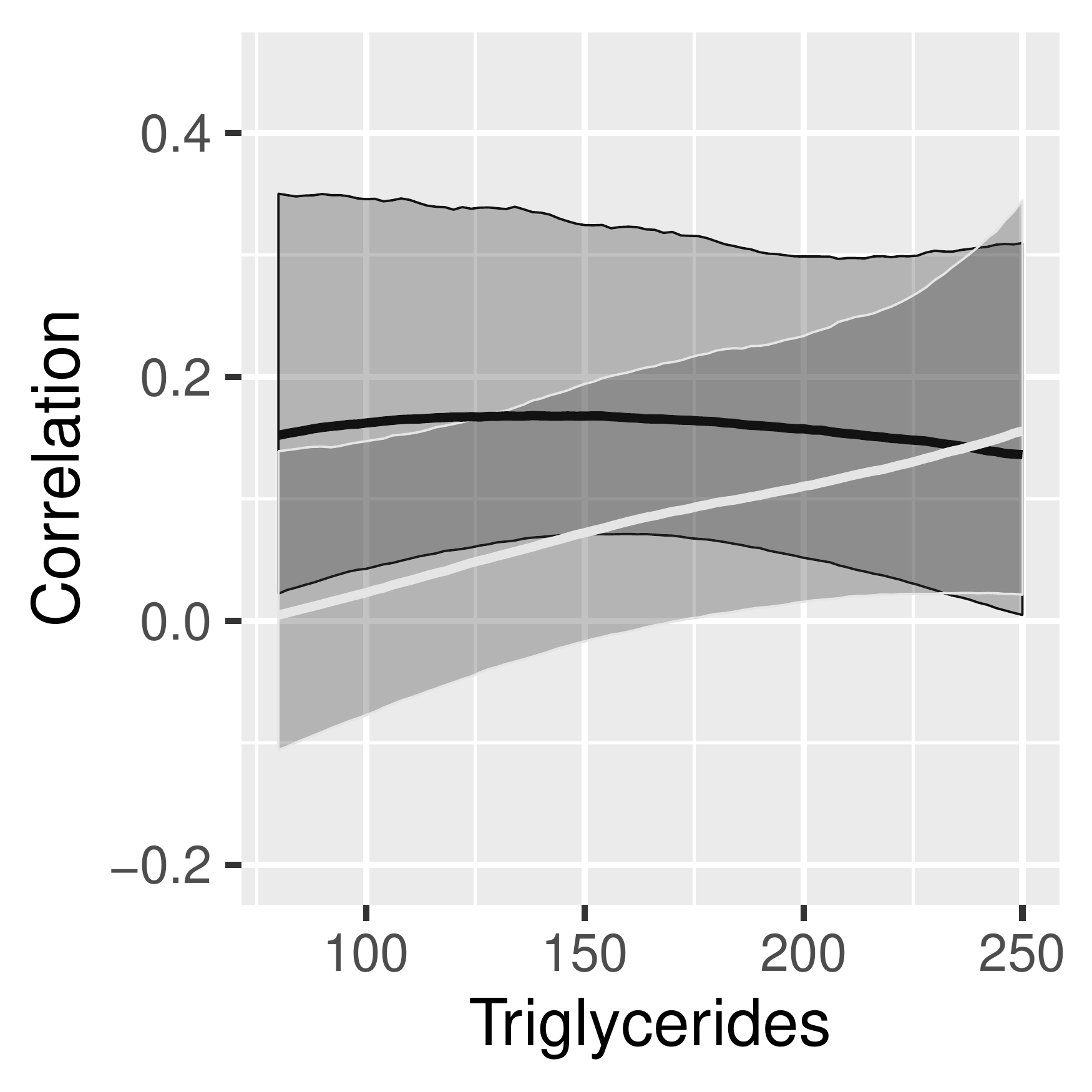}
} 
&
{
    \includegraphics[scale=0.3]{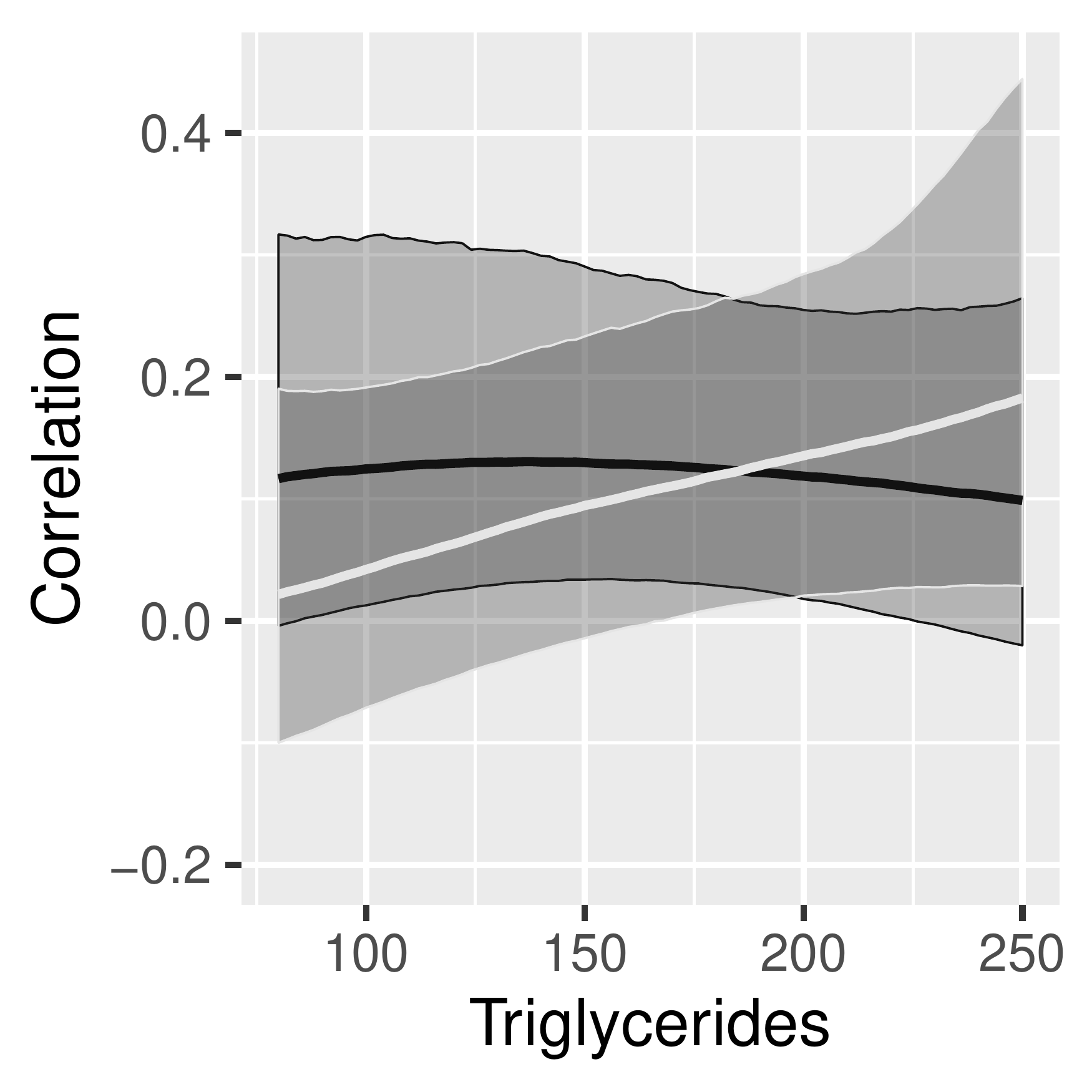}
}
&
{
    \includegraphics[scale=0.3]{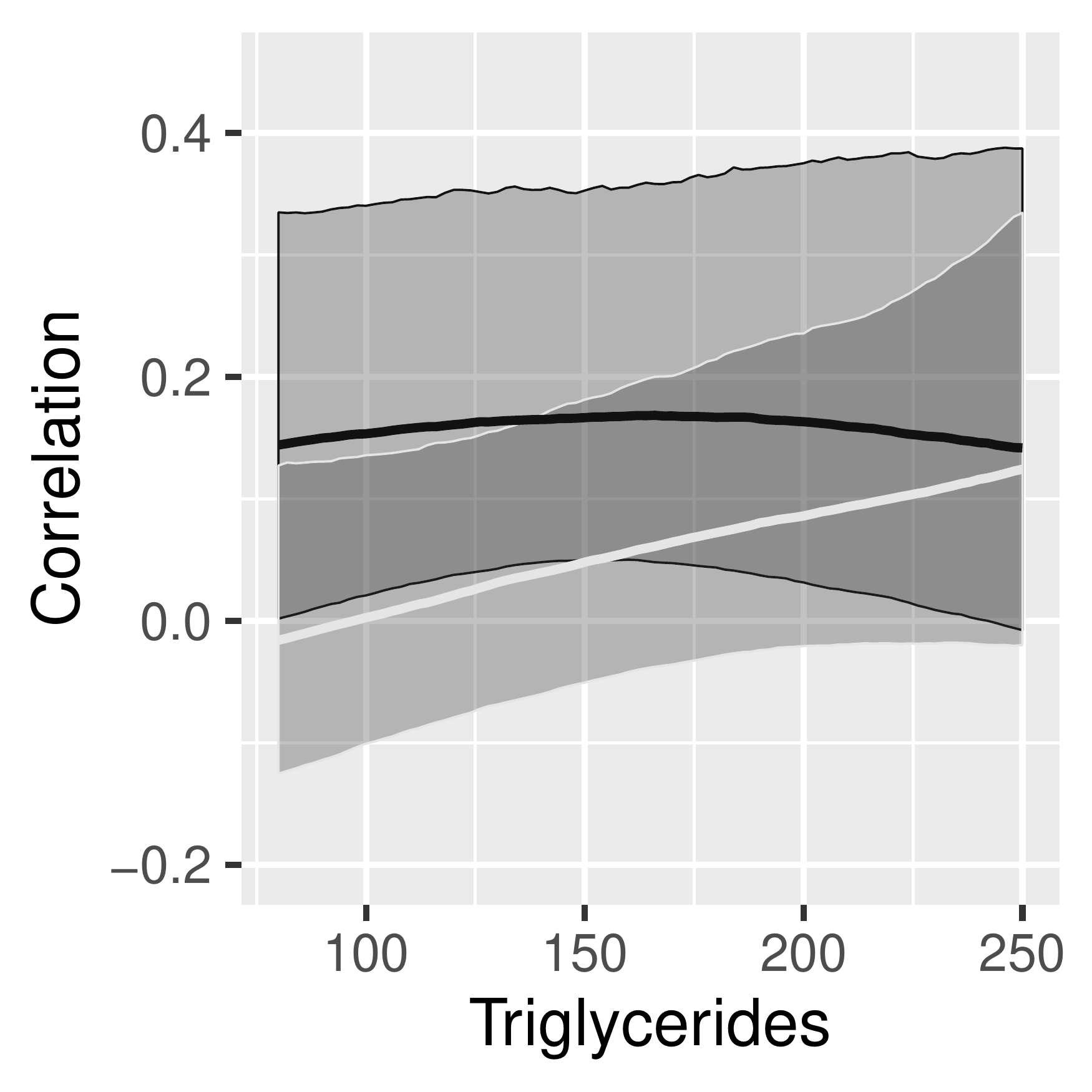}
}
&
{
    \includegraphics[scale=0.3]{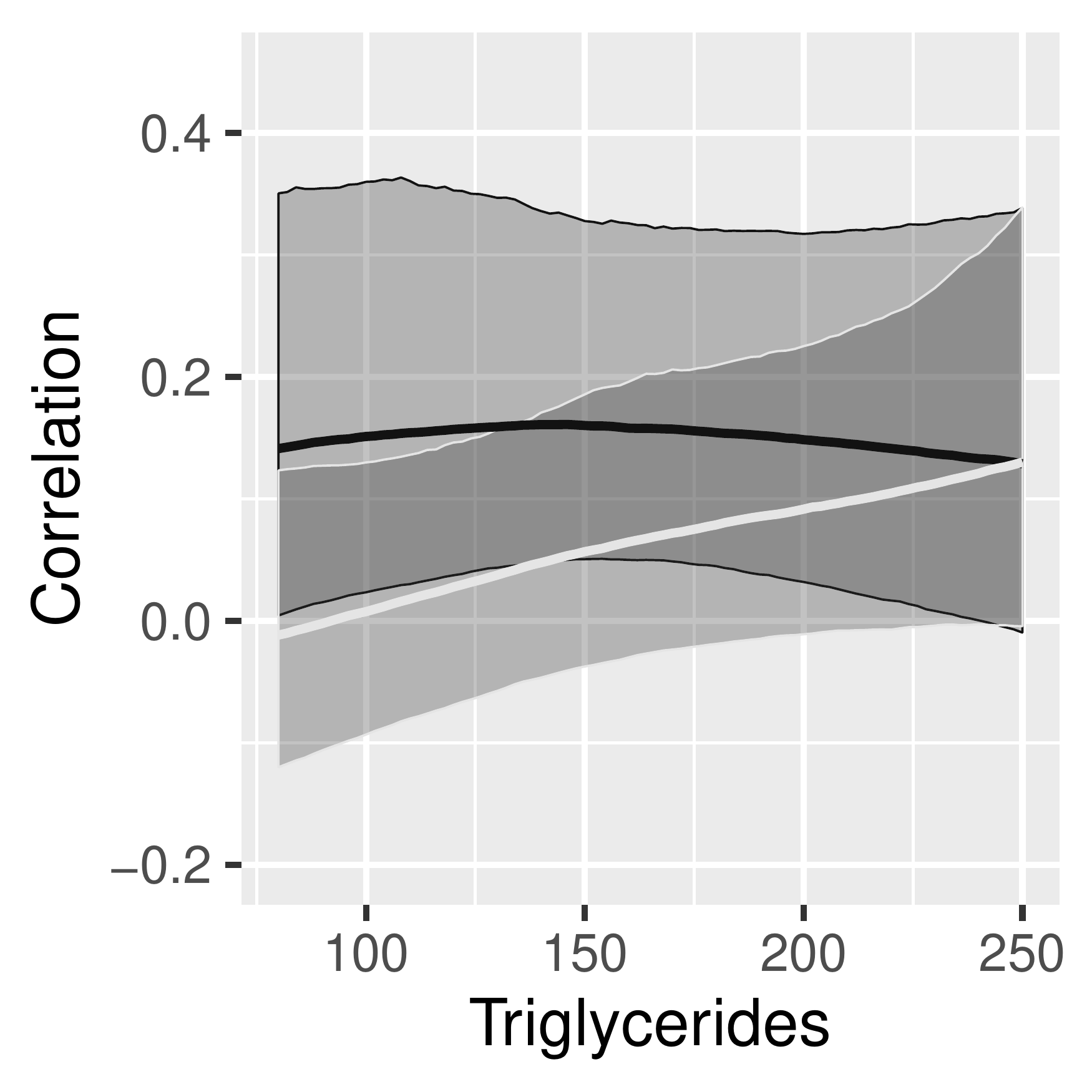}
}
\end{tabular}
}
\caption{\label{app:curveKendallsTauE1E2BMI32}\small{Mexico City Diabetes Study Data: section plot of conditional Kendall's tau  between CIMT and  PG2-H (dark grey)  and between CIMT and  Hb1Ac (light grey) when BMI  equals 32. Panels display the estimate of  Kendall's tau for varying values of triglycerides, sex, and age category for normoglycemic participants: the continuous line is the estimate (dark grey for Kendall's tau  between CIMT and  PG2-H and light grey for Kendall's tau  between CIMT and  Hb1Ac) while the grey region is a 90\% credible band. The y-axis shows values for Kendall's tau while x-axis shows values of triglycerides.}}
\end{figure}

Figures \ref{app:curveKendallsTauE1E2BMI22} - \ref{app:curveKendallsTauE1E2BMI32} 
display conditional Kendall's tau  between CIMT and PG2-H (dark grey) and between CIMT and HbA1c (light grey) for the normoglycemic participants in the MCDS for three different levels of BMI, namely $22, 27$ and $32$, respectively. We consider the values  22, 27, and 32 of BMI as indicative of individuals with normal weight (defined as BMI between $18.5-24.9$), overweight (defined as BMI between  $25-29.9$), and obese  (BMI $30$ or above), respectively.  Panels display conditional Kendall's tau estimate  (pointwise posterior predictive median) and 90\% credible bands for triglycerides in a range between $(80,250)$, males, females, and the four categories of age. From these figures, it is clear that the association structure, described by Kendall's tau, varies for different value of the covariates (tryglicerides, sex and age) in shape and strength. In the three figures, the association between PG2-H and CIMT for the normoglycemic patients is a slightly  decreasing  positive function of the triglycerides levels for both, male and female, through  all age categories. Although the association is relatively small it is  stronger for higher levels of BMI. The association structure between HbA1c  and CIMT shows to be a roughly constant function of triglycerides for both male and female through all age categories, when  BMI equals  22, whereas these associations become  increasing functions of the triglycerides as BMI increases. Although appealing, the higher values of Kendall's tau between HbA1c  and CIMT  obtained for higher values of triglycerides do not seem to differ greatly from the corresponding Kendall's tau between PG2-H  and CIMT.

In order to gain  better understanding  regarding which hyperglycemic marker would provide more evidence of association with the cardiovascular risk marker for varying values of the predictors, we compute the test statistic of the  nonparametric Kendall's tau correlation test. For a large sample size, the predictor-dependent test statistic $S(\tau(\bx)) = \tau(\bx)/\sqrt{\frac{2(2n+5)}{9n(n-1)}}$, where $n$ is the sample size (here $n=246$), follows approximately a standard normal distribution. For each covariate, we compute the proportion   $P(\bx) = \frac{1}{M}\sum_{m=1}^M\mathbbm{1}(\vert S(\tau(\bx)^{(m)})\vert > z_p)$, where $M$ denotes the number of MCMC scans,  $\tau(\bx)^{(m)}$ denotes the $m$--th  posterior predictive conditional Kendall's tau, $\vert\cdot \vert$ denotes the absolute value function,  $\mathbbm{1}(\cdot)$ denotes the indicator function, and $z_p$ denotes the $p$--th quantile of the standard normal distribution, specifically the $0.975$ quantile.  When computing the  proportion, $P(\bx)$, for one of the hyperglycemic markers and CIMT, the greater the proportion, the more evidence to support that hyperglycemic marker as an additional guidance for cardiovascular risk. 

\begin{figure}
\centering
\scalebox{0.55}{
\begin{tabular}{ccccc}
&\large{$age < 55$} & \large{$55 \leq age < 60$} & \large{$60 \leq age < 65$}  & \large{$age \geq 65$} \\
\rot{\hspace{1.8cm} \large{Male}} &
{
    \includegraphics[scale=0.3]{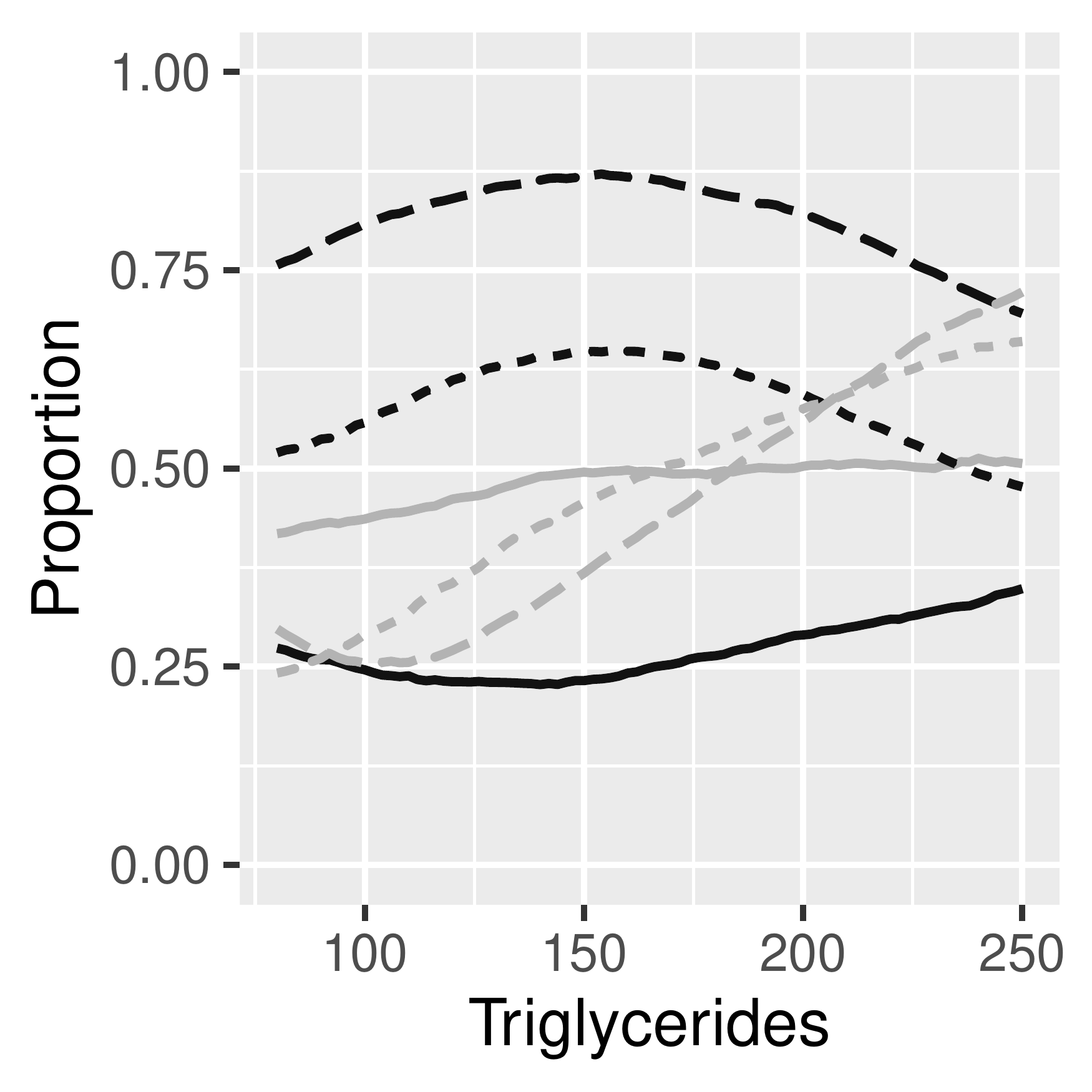}
}
&
{
    \includegraphics[scale=0.3]{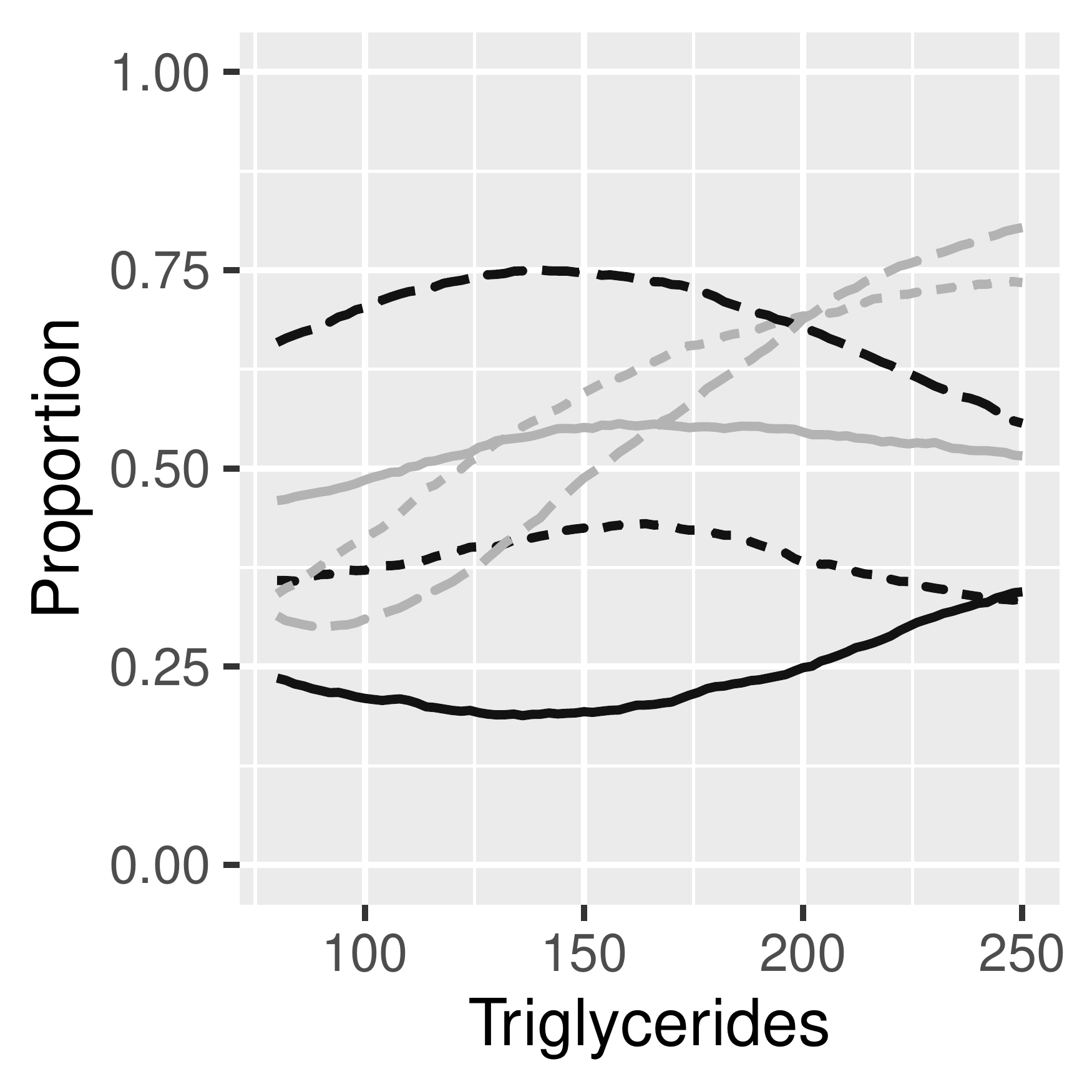}
}
&
{
    \includegraphics[scale=0.3]{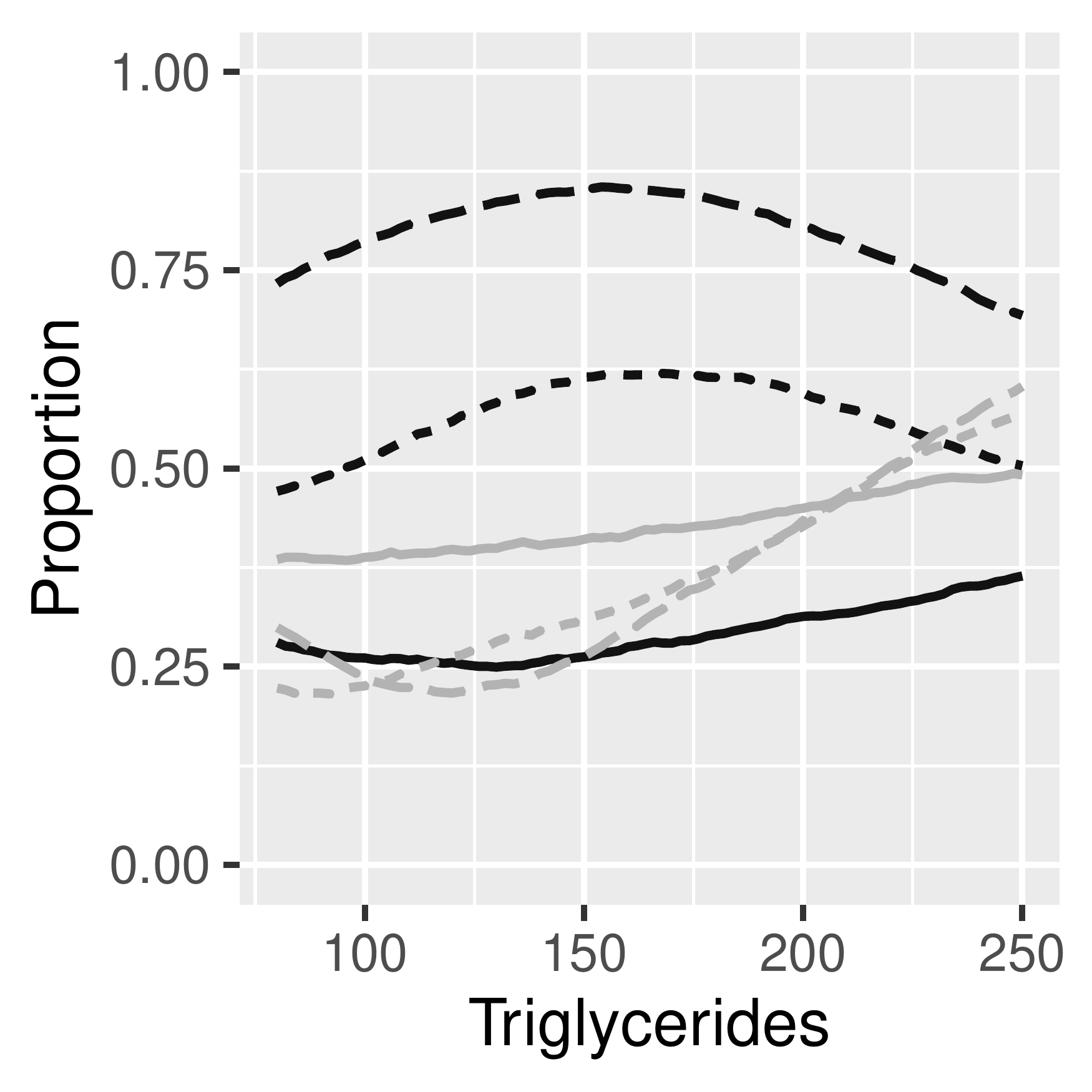}
}
&
{
    \includegraphics[scale=0.3]{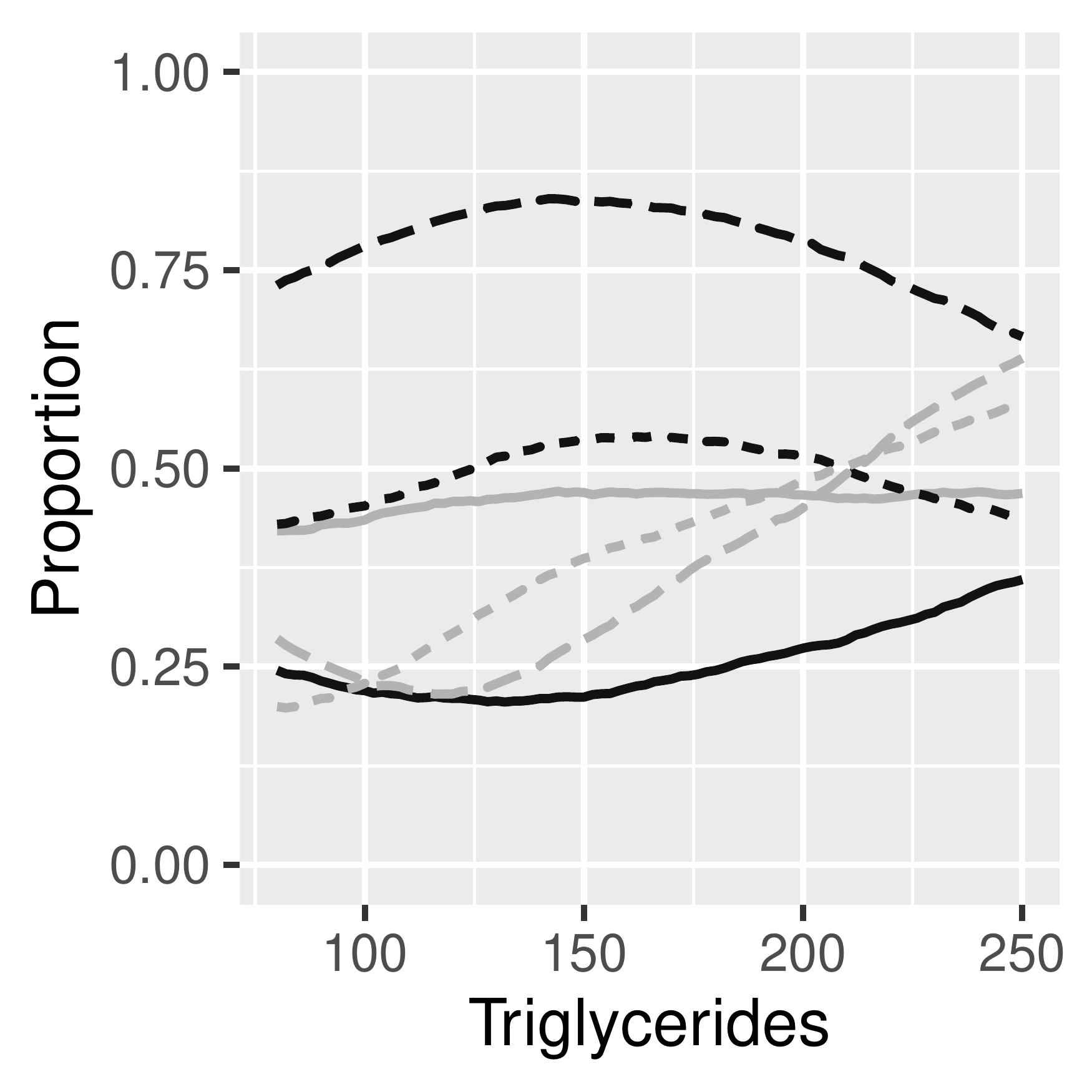}
} 
\\
\rot{\hspace{1.8cm}  \large{Female}}
&
{
    \includegraphics[scale=0.3]{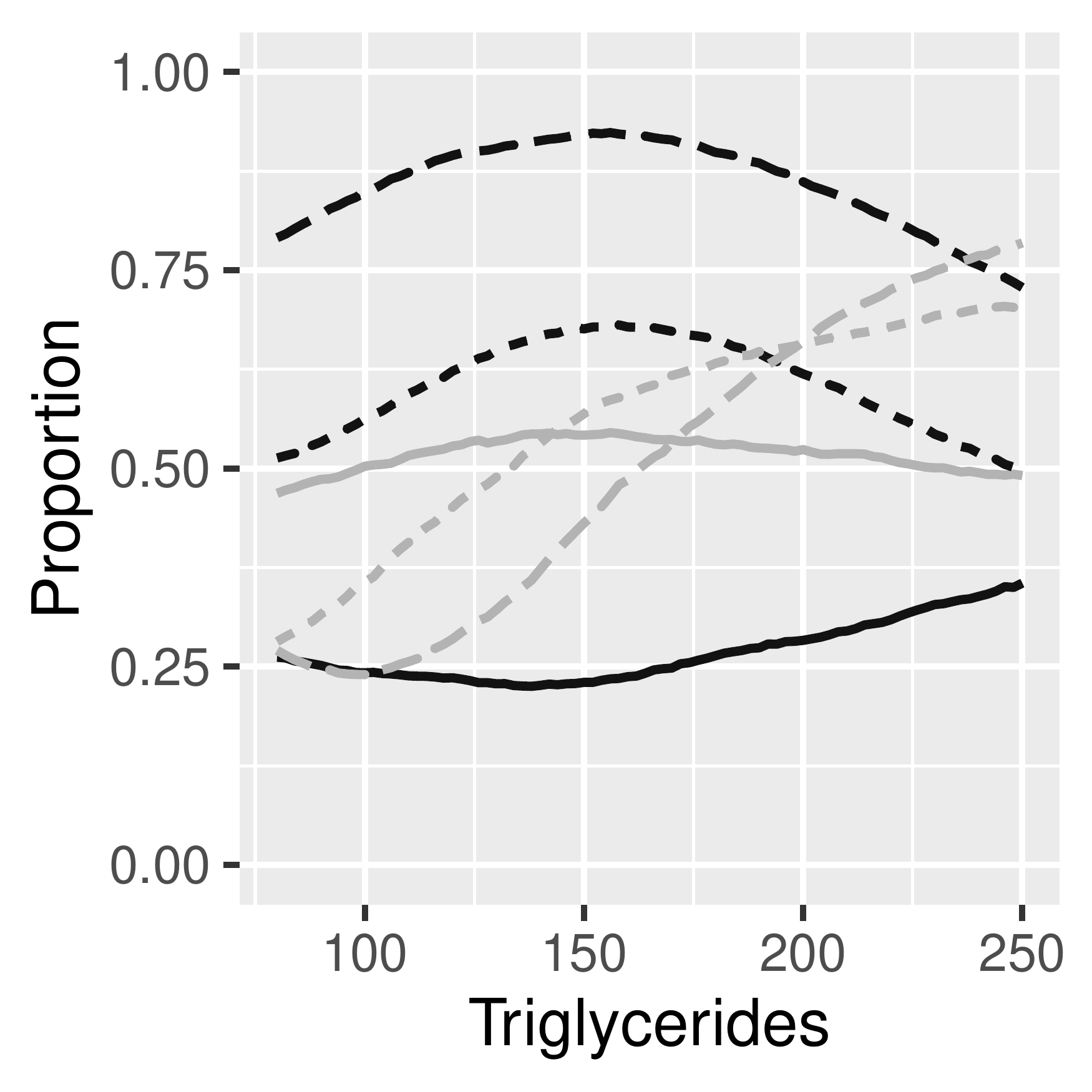}
} 
&
{
    \includegraphics[scale=0.3]{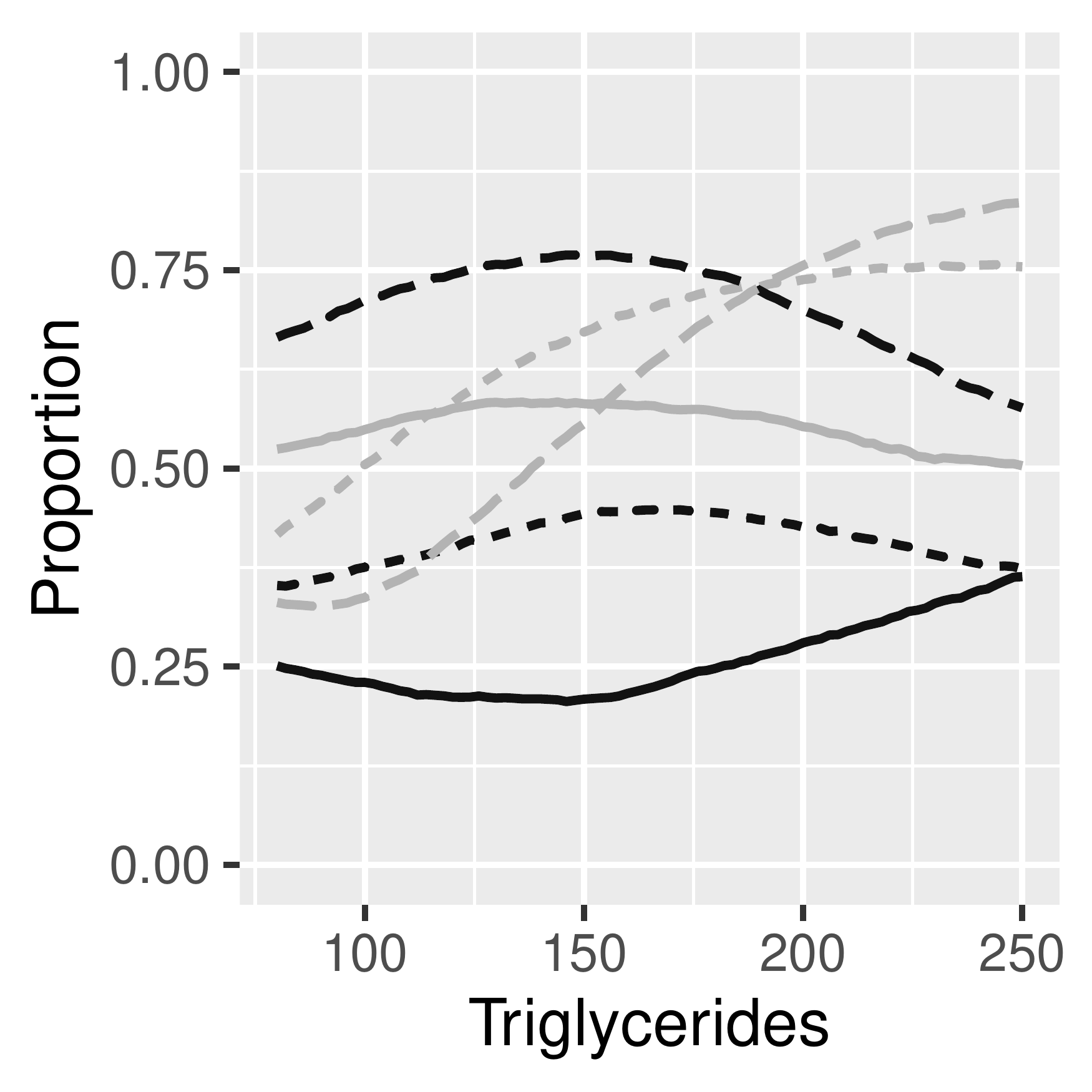}
}
&
{
    \includegraphics[scale=0.3]{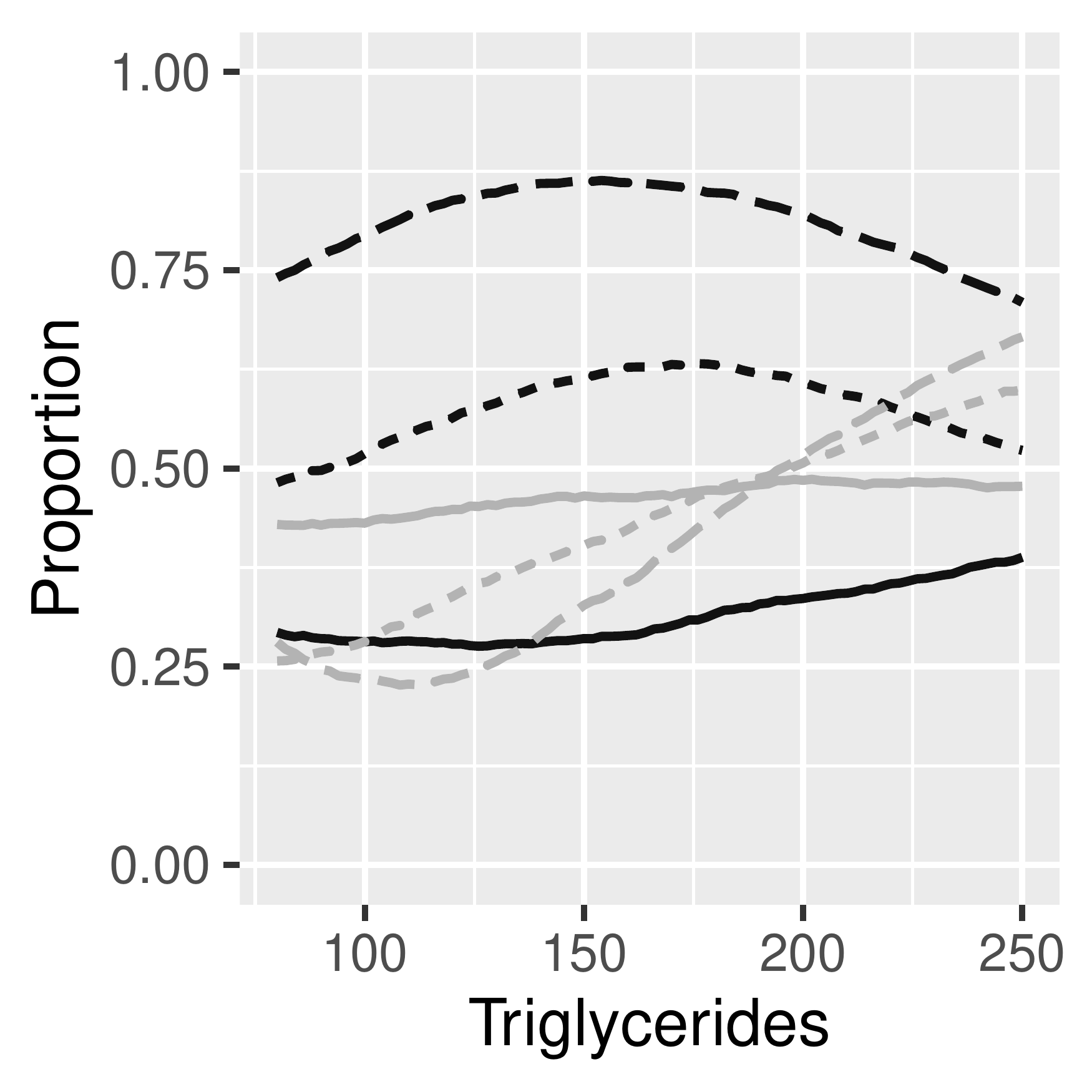}
}
&
{
    \includegraphics[scale=0.3]{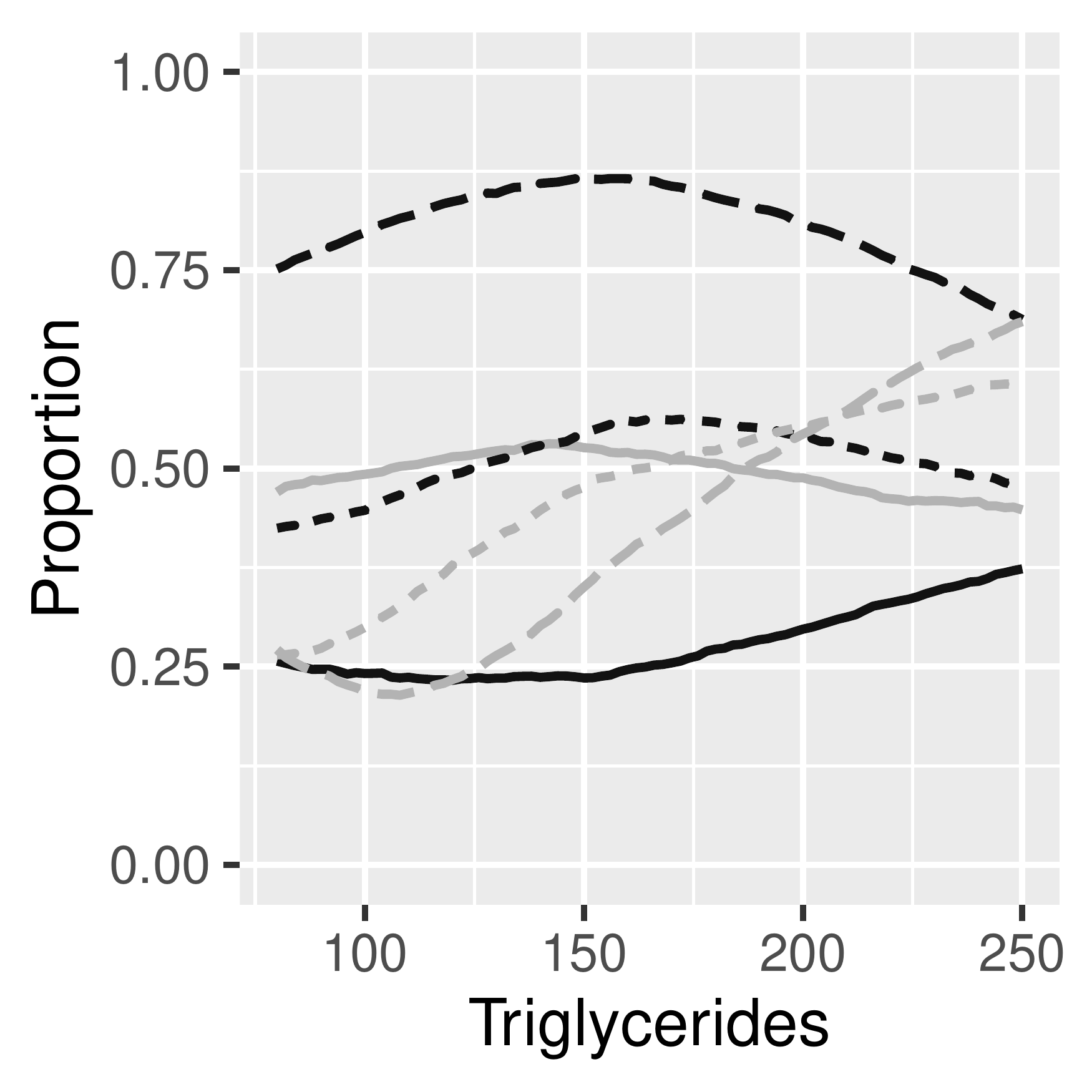}
}
\end{tabular}
}
\caption{\label{app:proportionSignificantTestStatisticReduced}\small{Mexico City Diabetes Study Data: proportion of test statistics that are greater than $z_{0.975}$  for Kendall's tau between   CIMT and PG2-H (dark grey) and between CIMT and HbA1c (light grey) for the normoglycemic participants in the MCDS for sex, all categories of age,   BMI equal to 22 (solid lines), BMI equal to 27 (dotted lines), and BMI equal to 32 (dashed lines).}}
\end{figure}

Figure  \ref{app:proportionSignificantTestStatisticReduced}  displays the proportions, $P(\bx)$, for Kendall's tau between   CIMT and PG2-H (dark grey) and between CIMT and HbA1c (light grey) for the normoglycemic participants in the MCDS for both male and female, all categories of age, and  BMI levels of 22, 27, and 32. Although the posterior predictive conditional Kendall's tau are generally small, the proportion of test statistics that are greater than $z_{0.975}$ are high for some covariate combinations. Particularly, the proportions for the test statistics of Kendall's tau between CIMT and PG2-H, for individuals with BMI equal to 32.

For individuals with BMI equal to 22, HbA1c gives more evidence of cardiovascular risk  than PG2-H, uniformly for every value of the considered predictors.  For individuals with BMI equal to 27, all age categories, except between 55 and 60 years, and both, male and female,  PG2-H  gives more evidence of cardiovascular risk for triglyceride levels lower than \~200, while for greater levels of triglycerides HbA1c is a better cardiovascular risk marker. Interestingly, for individuals aged between 55 and 60 years, HbA1c provides more evidence of cardiovascular risk  than PG2-H, for every triglyceride level. Finally, for individuals with BMI equal to 32, all age categories, except between 55 and 60 years, PG2-H   gives more evidence of cardiovascular risk uniformly across triglyceride levels, while for individuals aged between 55 and 60, HbA1c  provides more evidence of cardiovascular risk for  individuals with high triglyceride levels (greater than \~200). Another interesting finding, is that the use of PG2-H  marker shows stronger evidence of association with CIMT as the BMI increases, uniformly across age categories and triglyceride levels, while this is observed  for HbA1c only for triglyceride levels greater than \~200.

Although appealing due to the ready availability  of the test, our findings indicate that the use of HbA1c as a cardiovascular risk marker is not uniformly better than a PG2-H test, across mediating covariates in the MCDS population. 

\section{Conclusion}
\label{sec:conc}

Understanding T2D and its relationship with other conditions is an important and complex task. Rates of T2D are increasing world wide and in particular, for the Mexican population. The assessment of cardiovascular risk through hyperglycemic markers is relevant. Although   HbA1c is an accessible marker, our findings show that it does not provide more evidence of cardiovascular risk than PG2-H, uniformly across covariates, in normoglycemic subjects of the MCDS.

In this article we have presented  a general and flexible model for describing the association structure between hyperglycemic markers and cardiovascular risk, controlled by triglycerides, BMI, age, and sex.   Our methodology is based on a  fully BNP approach for modeling conditional copula densities. As a byproduct the conditional  association structure described by the conditional Kendall's tau is obtained. Therefore,  a direct understanding of this association, adjusted by  known risk mediators, is feasible without the need to define thresholds in  either, the CIMT,  HbA1c or PG2-H.

\bibliographystyle{agsm}
\bibliography{Bibliography}

\end{document}